\documentclass[trackchanges,twocolumn,  twocolappendix]{aastex701}

\usepackage{graphicx, bm, amssymb, xcolor}
\usepackage{verbatim}
\usepackage[all]{hypcap}
\usepackage{xspace}
\usepackage{multirow,bigdelim}
\usepackage[fleqn]{amsmath}

\newcommand{\UGent}{Sterrenkundig Observatorium, Universiteit Gent, Krijgslaan 281 S9, B-9000 Gent, Belgium}

\newcommand{\MPIA}{Max-Planck-Institut f\"{u}r Astronomie, K\"{o}nigstuhl 17, D-69117, Heidelberg, Germany}

\newcommand{\UCSD}{Department of Astronomy \& Astrophysics, University of California, San Diego, 9500 Gilman Dr., La Jolla, CA 92093, USA}

\newcommand{\OSU}{Department of Astronomy, The Ohio State University, 140 West 18th Avenue, Columbus, OH 43210, USA}

\newcommand{\CCAPP}{Center for Cosmology and Astroparticle Physics (CCAPP), 191 West Woodruff Avenue, Columbus, OH 43210, USA}

\newcommand{\ARI}{Astronomisches Rechen-Institut, Zentrum f\"{u}r Astronomie der Universit\"{a}t Heidelberg, M\"{o}nchhofstr. 12-14, D-69120 Heidelberg, Germany}

\newcommand{\UWyoming}{Department of Physics and Astronomy, University of Wyoming, Laramie, WY 82071, USA}

\newcommand{\ITA}{Universit\"{a}t Heidelberg, Zentrum f\"{u}r Astronomie, Institut f\"{u}r Theoretische Astrophysik, Albert-Ueberle-Str 2, D-69120 Heidelberg, Germany}

\newcommand{\ESO}{European Southern Observatory, Karl-Schwarzschild Stra{\ss}e 2, D-85748 Garching bei M\"{u}nchen, Germany}

\newcommand{\OAN}{Observatorio Astron{\'o}mico Nacional (IGN), C/ Alfonso XII 3, E-28014 Madrid, Spain}

\newcommand{\kipac}{Kavli Institute for Particle Astrophysics \& Cosmology (KIPAC), Stanford University, CA 94305, USA}

\newcommand{\cdu}{Center for Decoding the Universe, Stanford University, CA 94305, USA}

\newcommand{\stanford}{Department of Physics, Stanford University, Stanford, CA 94305, USA}

\newcommand{\Maryland}{Department of Astronomy, University of Maryland, 4296 Stadium Drive, College Park, MD 20742, USA}

\newcommand{\NRAOABQ}{National Radio Astronomy Observatory, 800 Bradbury SE, Suite 235, Albuquerque, NM 87106 USA}

\newcommand{\ANU}{Research School of Astronomy and Astrophysics, Australian National University, Canberra, ACT 2611, Australia}

\newcommand{\Whitman}{Whitman College, 345 Boyer Avenue, Walla Walla, WA 99362, USA}

\newcommand{\JBCA}{UK ALMA Regional Centre Node, Jodrell Bank Centre for Astrophysics, Department of Physics and Astronomy, The University of Manchester, Oxford Road, Manchester M13 9PL, UK}

\newcommand{\ASIAA}{Institute of Astronomy and Astrophysics, Academia Sinica, No. 1, Sec. 4, Roosevelt Road, Taipei 106319, Taiwan}

\newcommand{\UToledo}{Ritter Astrophysical Research Center, University of Toledo, Toledo, OH 43606, USA}

\newcommand{\NAOJ}{National Astronomical Observatory of Japan, 2-21-1 Osawa, Mitaka, Tokyo 181-8588, Japan}
\newcommand{\MPIfR}{Max Planck Institute for Radio Astronomy, Auf dem Hügel 69, 53121 Bonn, Germany}

\newcommand{\AlgomaU}{Faculty of Computer Science \& Technology, Algoma University, Sault Ste. Marie, Ontario, P6A 2G3, Canada}
\shorttitle{High CO-to-PAH Ratios in PHANGS Galaxies}
\shortauthors{Kim et al.}

\begin{document}

\title{Localized Deviations from the CO–PAH Relation in PHANGS–JWST Galaxies: Faint PAH Emission or Elevated CO Emissivity?}

\author[orcid=0000-0002-0432-6847]{Jaeyeon Kim}\thanks{Kavli Postdoctoral Fellow}
\affiliation{\kipac}
\email[show]{jyeonkim@stanford.edu} 

\author[0000-0002-2545-1700]{Adam~K.~Leroy}
\affiliation{\OSU}
\affiliation{\CCAPP}
\email{leroy.42@osu.edu}

\author[0000-0002-4378-8534]{Karin~Sandstrom}
\affiliation{\UCSD}
\email{kmsandstrom@ucsd.edu}

\author[orcid=0000-0002-0012-2142]{Sharon E. Meidt}
\affiliation{\UGent}
\email{sharon.vanderwel@ugent.be}

\author[0000-0003-4209-1599]{Yu-Hsuan~Teng}
\affiliation{\Maryland}
\email{yhteng@umd.edu}

\author[0000-0002-0472-1011]{Miguel Querejeta}
\affiliation{\OAN}
\email{m.querejeta@oan.es}

\author[0000-0002-3933-7677]
{Eva~Schinnerer}\affiliation{\MPIA}
\email{schinner@mpia.de}

\author[0000-0002-7633-3376]{Susan~E.~Clark}
\affiliation{\stanford}
\affiliation{\kipac}
\email{seclark1@stanford.edu}

\author[0000-0001-8241-7704]{Ryan~Chown}
\affiliation{\AlgomaU}
\affiliation{\OSU}
\email{ryan.chown@algomau.ca}

\author[0000-0001-6708-1317]{Simon C.~O.\ Glover}
\affiliation{\ITA}
\email{glover@uni-heidelberg.de}

\author[0000-0002-5782-9093]{Daniel~A.~Dale}
\affiliation{Department of Physics and Astronomy, University of Wyoming, Laramie, WY 82071, USA}
\email{ddale@uwyo.edu}

\author[0000-0003-4974-3481]{Dalya Baron}
\affiliation{\kipac}
\affiliation{\cdu}
\email{dalbaron@stanford.edu}

\author[0000-0002-9183-8102]{Jessica Sutter}
\email[]{sutterjs@whitman.edu}
\affiliation{\Whitman}

\author[orcid=0000-0003-0410-4504]{Ashley T. Barnes}
\affiliation{\ESO}
\email{ashleybarnes.astro@gmail.com}

\author[0000-0002-8760-6157]
{Jakob den Brok}\affiliation{\MPIA}
\email{jadenbrok@mpia.de}

\author[0000-0003-0085-4623]{Rupali~Chandar}
\affiliation{\UToledo}
\email{Rupali.Chandar@utoledo.edu}

\author[orcid=0000-0003-2551-7148]{I-Da Chiang}
\affiliation{\ASIAA}
\email{ida.chiang.tw@gmail.com}

\author[orcid=0000-0002-4755-118X]{Oleg~V.~Egorov}
\affiliation{\ARI}
\email{oleg.egorov@uni-heidelberg.de}

\author[0000-0002-3247-5321]{Kathryn~Grasha}
\altaffiliation{ARC DECRA Fellow}
\affiliation{\ANU}   
\email{kathryn.grasha@anu.edu.au}

\author[0000-0002-0560-3172]{Ralf S.\ Klessen}
\affiliation{\ITA}
\affiliation{Universit\"{a}t Heidelberg, Interdisziplin\"{a}res Zentrum f\"{u}r Wissenschaftliches Rechnen, Im Neuenheimer Feld 225, 69120 Heidelberg, Germany}
\email{klessen@uni-heidelberg.de}

\author[orcid=0000-0001-6551-3091]{Kathryn Kreckel}
\affiliation{\ARI}
\email{kathryn.kreckel@uni-heidelberg.de}

\author[0000-0001-9605-780X]{Eric W. Koch}
\affiliation{\NRAOABQ}
\email{koch.eric.w@gmail.com}

\author[0009-0001-5949-1524]{Hannah Koziol}
\affiliation{\UCSD}
\email{hkoziol@ucsd.edu}

\author[0000-0001-9793-6400]{Lukas~Neumann}
\affiliation{\ESO}
\email{lukas.neumann@eso.org}

\author[0000-0002-1370-6964]{Hsi-An Pan}
\affiliation{Department of Physics, Tamkang University, No.151, Yingzhuan Road, Tamsui District, New Taipei City 251301, Taiwan} 
\email{hapan@gms.tku.edu.tw}

\author[orcid=0000-0002-9333-387X]{Sophia K. Stuber}
\affiliation{\NAOJ}
\affiliation{\MPIfR}
\altaffiliation{JSPS Postdoctoral Fellow}
\email{astro@sophiastuber.de}

\author[0009-0005-8923-558X]{Tony D. Weinbeck}
\affiliation{\UWyoming}
\email{weinbeck@alum.mit.edu}

\author[orcid=0000-0002-0012-2142]{Thomas G. Williams}
\affiliation{\JBCA}
\email{thomas.williams-4@manchester.ac.uk}

\suppressAffiliations

\begin{abstract} 
 Polycyclic aromatic hydrocarbon (PAH) emission is widely used to trace the distribution of molecular gas in the interstellar medium, exhibiting a tight correlation with CO(2–1) emission across nearby galaxies. Using PHANGS-JWST and PHANGS-ALMA data, we identify localized regions where this correlation fails, with CO flux exceeding that predicted from 7.7$\mu$m PAH emission by more than an order of magnitude. These outlier regions are found in 20 out of 70 galaxies and are located in galaxy centers and bars, without signs of massive star formation. We explore two scenarios to explain the elevated CO-to-PAH ratios, which can either be due to suppressed PAH emission or enhanced CO emissivity. We examine PAH emission in other bands (3.3$\mu$m and 11.3$\mu$m) and the dust continuum dominated bands (10$\mu$m and 21$\mu$m), finding consistently high CO-to-PAH (or CO-to-dust continuum) emission ratios, suggesting that 7.7\,$\mu$m PAH emission is not particularly suppressed. In some outlier regions, PAH sizes and spectral energy distribution of the radiation differ slightly from nearby control regions with normal CO-to-PAH ratios, though without a consistent trend. We find that the outlier regions show higher CO velocity dispersions ($\Delta v_{\mathrm{CO}}$). This increase in $\Delta v_{\mathrm{CO}}$ lowers CO optical depth and raises its emissivity for a given gas mass. Our results favor a scenario where shear along the bar lanes and shocks at the bar ends elevate CO emissivity, leading to the breakdown of the CO–PAH correlation. Future JWST spectroscopy  and  deep ALMA observations of CO isotopologues will provide critical tests of this scenario.
\end{abstract}



\section{Introduction}\label{sec:intro}

Emission from polycyclic aromatic hydrocarbons (PAHs) captured with JWST reveals a complex and widespread network of interstellar medium (ISM) structures throughout galaxy disks \citep{lee23, thilker23, fisher24, williams24}. In the diffuse Milky Way ISM, the fraction of cold neutral gas with respect to the total gas strongly correlates with PAH abundance, suggesting that PAHs are destroyed in warm gas while PAH growth and survival is facilitated in well-shielded, moderately dense gas \citep{galliano18, hensley23}. In molecular gas-dominated galaxies, studies have shown a tight correspondence of PAH emission with CO emission as well as with H$\alpha$ at scales ranging from a few kpc down to 100\,pc \citep{regan06,chown21, gao22,  leroy23, whitcomb23, chown24, shim25}.

The tight correlation observed between CO and PAH emission in molecular gas-dominated galaxies can be understood within the standard framework of dust in galaxies, in which dust and PAH grains are well mixed within the gaseous phase of the ISM \citep{draine07, galliano18, hensley23}. PAHs are stochastically excited by ultraviolet and optical photons, allowing them to emit efficiently in the mid-infrared. As a result, the distribution of mid-IR emission is expected, to first order, to closely resemble the distribution of the cold ISM \citep{leroy23, chown24}. Indeed, PAH emission has even been used to probe the ISM in \textsc{Hi}-dominated regions with gas surface densities below $\sim 7\,M_{\odot}\,\mathrm{pc}^{-2}$ \citep{sandstrom23}.

The excitation of PAHs also depends on the intensity of the radiation field, leading to brighter PAH emission near star-forming regions \citep{pathak24}. \citet{leroy23} demonstrated that mid-IR emission can be described as a linear combination of scaled CO and H$\alpha$ fluxes. However, strong ionizing radiation from young massive stars also destroys PAHs, as evidenced by suppressed PAH emission at 7.7 and 11.3$\mu$m relative to the hot dust continuum at 21$\mu$m in \textsc{Hii} regions \citep{gordon08, chastenet19, chastenet23, egorov23, Egorov2025, sutter24}.

Recently, \citet{chown24} have characterized the relationship between $^{12}$CO($J{=}2{-}1$; hereafter CO) and PAH emission across 70 galaxies from the Physics at High Angular Resolution in Nearby Galaxies Survey (PHANGS). By combining Atacama Large Millimeter/sub-millimeter Array (ALMA) CO maps and JWST near- and mid-infrared (near-IR and mid-IR) imaging, \citet{chown24} have found a relatively small scatter ($\sim0.2{-}0.5\,\mathrm{dex}$) in the observed CO-to-PAH ratios at $\sim$100\,pc-scale, where CO shows a strong and approximately linear correlation with 7.7$\mu$m PAH emission, one of the common PAH tracers accessible with JWST. Similarly linear trends were observed at 11.3$\mu$m and 3.3$\mu$m for a subsample of 19 galaxies from the PHANGS–JWST Cycle~1 survey \citep{lee23}.
 
The growing evidence of PAH closely tracing CO has led to the suggestion that PAH emission can be used as a tracer of the cold ISM, not only for high–resolution studies of nearby galaxies but also out to high redshift \citep{cortzen19,shivaei24,chown24,shim25}. This prospect is particularly exciting because JWST is now routinely enabling high–resolution, high–sensitivity PAH imaging. Looking ahead, Spectro-Photometer for the History of the Universe, Epoch of Reionization and Ices Explorer satellite (SPHEREx; \citealp{dore18}) will deliver full–sky spectral PAH maps, opening the door to uniform cold ISM mapping across all nearby galaxies. However, for these applications to be reliable, we need to identify where the relation between PAH and cold gas breaks down, beyond classical \textsc{Hii} regions, for example in AGN-dominated centers and zones of strong shear or shocks.

In this Letter, we examine regions that deviates significantly from the CO–7.7$\mu$m PAH relation as calibrated by \citet{chown24} across the same PHANGS galaxies. We then assess the local environment, galactic dynamics, and excitation conditions of these outliers to isolate the physical processes that drive these deviations from the global correlation.

\section{Observational Data}\label{sec:data}
We focus on the same 70 galaxies from \citet{chown24} for which scaling relations between CO and 7.7$\mu$m PAH emission were derived using high-resolution JWST imaging with the F770W filter and ALMA CO observations. These targets are from PHANGS–JWST Cycle~1 (GO~2107; PI: Lee; \citealp{lee23, williams24}) and Cycle~2 (GO~3707; PI: Leroy; \citealp{chown24}). Most of these galaxies lie on the star-forming main sequence, with stellar masses ranging from $\log(M_{*}/M_{\odot}) = 9$ to 11.5 and at distances of $D=5{-}25$\,Mpc.

\subsection{JWST mid-IR imaging of PAH emission}\label{ssec:data_jwst}
For 19 galaxies from the PHANGS-JWST Cycle~1 program, we use JWST NIRCam and MIRI imaging data observed with F335M, F770W, F1000W, F1130W, and F2100W filters. The Cycle~1 data can be obtained using this link\footnote{\url{https://archive.stsci.edu/hlsp/phangs} or \citet{doi}}. The remaining 51 galaxies are from the PHANGS-JWST Cycle~2 program described in \citet{chown24}, with limited coverage in filters. We use two mid-IR bands of F770W and F2100W. 

The F335M, F770W and F1130W filters capture strong PAH emission features at 3.3, 7.7 and 11.3$\mu$m. The F1000W (centered at 10$\mu$m) is expected to be dominated by dust continuum \citep{whitcomb23}, though its morphology and intensity often track CO and H$\alpha$, similar to the PAH bands \citep{leroy23}. The F2100W at 21$\mu$m captures dust continuum emission, including contributions from grains in thermal equilibrium and heated by the interstellar radiation field \citep{draine07, kennicutt12, leroy12, belfiore23_jwst, hassani23, leroy23, pathak24}. The resolution of F335M, F770W, F1000W, F1130W, F2100W imaging is 0\farcs11, 0\farcs27, 0\farcs33, 0\farcs38, and 0\farcs67, respectively\footnote{Point spread functions for \href{https://jwst-docs.stsci.edu/jwst-near-infrared-camera/nircam-performance/nircam-point-spread-functions\#NIRCamPointSpreadFunctions-PSFFWHM}{NIRCAM} and \href{https://jwst-docs.stsci.edu/jwst-mid-infrared-instrument/miri-performance/miri-point-spread-functions\#gsc.tab=0}{MIRI}}, where the resolution of F2100W roughly translates into a physical scale of $15-85$\,pc across the range of distances of our galaxy sample ($5{-}25$\,Mpc). Both of the Cycle~1 and Cycle~2 data have been reduced using the PHANGS-JWST pipeline (\texttt{pjpipe}\footnote{\url{https://pjpipe.readthedocs.io/en/latest/}}). For comprehensive descriptions of the survey and the data reduction process, we refer readers to \citet{lee23}, \citet{williams24}, and \citet{chown24}.

The F335M and F770W bands include contributions from the stellar continuum. For F335M, we use continuum-subtracted maps from H.\ Koziol et al.\ (in prep.), where the stellar continuum is estimated using F300M and F360M, based on the method of \citet{sandstrom23_33pah}. As for F770W, we used  stellar continuum-subtracted F770W maps adopted in \citet{chown24}, obtained by subtracting the flux in F200W (Cycle~1) or F300M (Cycle~2) scaled by a factor determined using the prescription of \citet{sutter24}. 

\subsection{ALMA \texorpdfstring{$^{12}$CO($J{=}2{-}1)$}~~imaging}\label{ssec:data_alma}
We use $^{12}$CO($J{=}2{-}1$) imaging from the PHANGS–ALMA survey, hereafter CO. Full descriptions of the sample, survey design, and image production process are provided in \citet{leroy21_survey} and \citet{leroy21_pipe}. The observations were conducted with the 12-m, 7-m, and total power arrays of ALMA. The CO maps have resolutions of $\sim$1\farcs0-1\farcs7, corresponding to physical scales of 25–150\,pc for the galaxies in our sample. We use the first public release of the moment-0 maps, generated using an inclusive signal-masking scheme (``broad'' masking; see \citealp{leroy21_pipe}), designed to capture extended and faint emission and encompass all emission from each galaxy at the native resolution. We also use maps of the CO effective line width ($\Delta v_{\mathrm{CO}}$) from the same public release, produced with the more restrictive “strict” mask, which includes only voxels highly likely to contain real emission.

\subsection{ALMA \texorpdfstring{$^{12}$CO($J{=}1{-}0)$}~~imaging}\label{ssec:co10}
If outliers with abnormally high CO/PAH ratios are driven by CO excitation, the CO line ratios ($^{12}$CO$(J{=}2{-}1)/^{12}$CO$(J{=}1{-}0)$) could increase. In order to test this, we use ALMA $^{12}$CO$(J{=}1{-}0)$ imaging from den Brok et al. (in preparation; 2022.1.01479.S) obtained using 12-m, 7-m, and total power antennas \citep[see also][]{denbrok25}. The observations were available for 14 of the 70 PHANGS galaxies. Among these 14 galaxies, 4 galaxies host significant outliers. 

\subsection{MUSE ionized gas emission}\label{ssec:data_muse}
For PHANGS-JWST Cycle~1 galaxies, we leverage optical integral field spectroscopy from the PHANGS–MUSE survey \citep{emsellem22} in order to look for signs of massive star formation. The PHANGS–MUSE survey observed all 19 of the JWST Cycle~1 galaxies with the Multi Unit Spectroscopic Explorer (MUSE) on the Very Large Telescope (VLT), covering the star-forming disks at a spatial resolution of $\sim$1\farcs0 (20–100\,pc) over 4800–9300\,\textup{\AA}, enabling measurements of key emission lines including H$\beta$ and H$\alpha$. Emission line fluxes were derived by simultaneously fitting the simple stellar population models to continuum and a Gaussian distribution to gas emission lines. We use H$\alpha$ maps from \citet{belfiore23}, which has been corrected for dust extinction using the Balmer decrement. As for the PHANGS-JWST Cycle~2 galaxies without MUSE coverage for the regions of interest, we adopt continuum-subtracted narrow-band H$\alpha$ imaging from Razza et al. (in preparation).

\begin{figure*}
\centering
\includegraphics[scale=0.5]{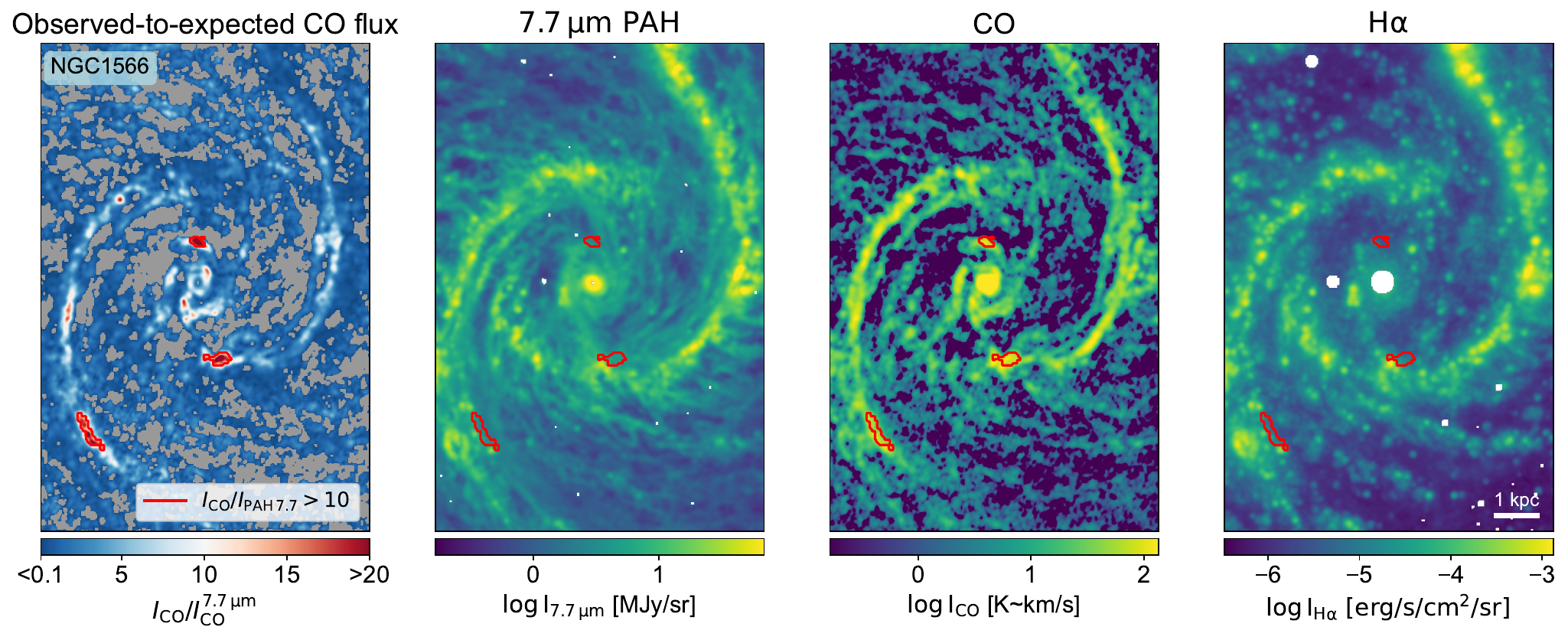}
\includegraphics[scale=0.5]{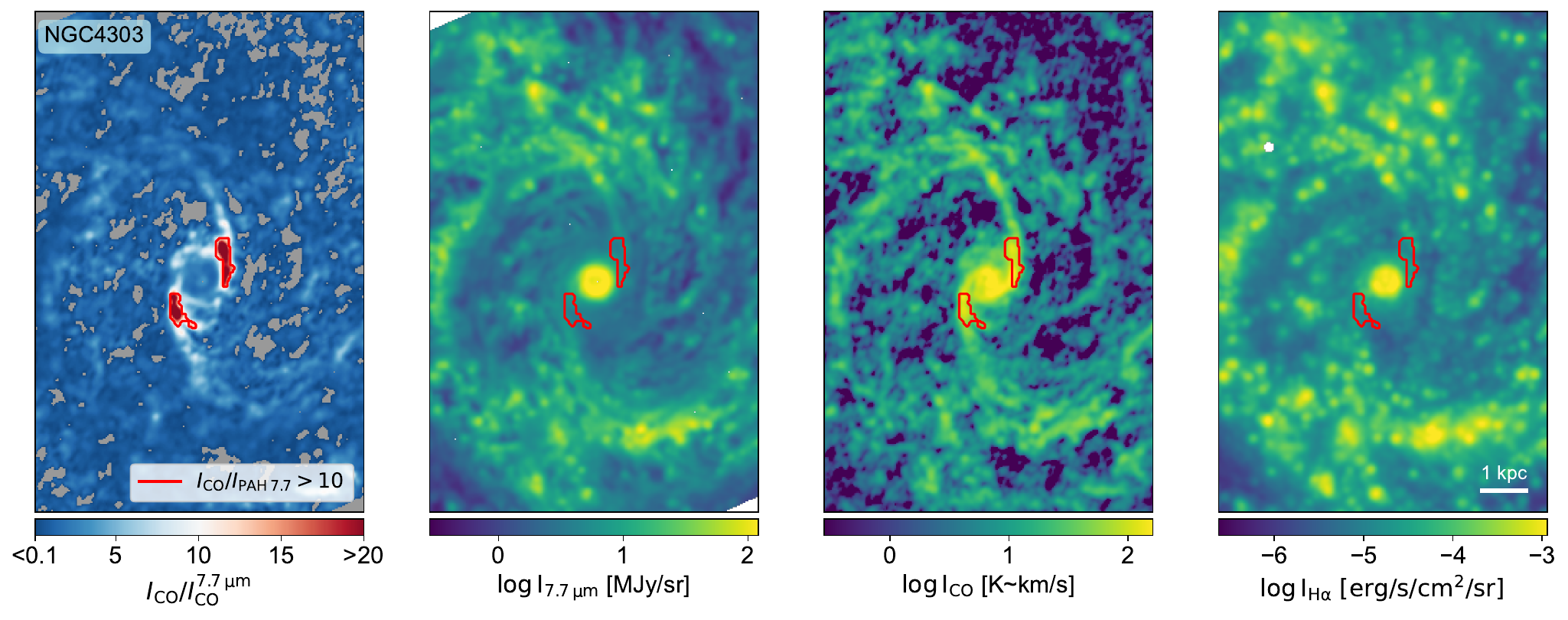}
\includegraphics[scale=0.5]{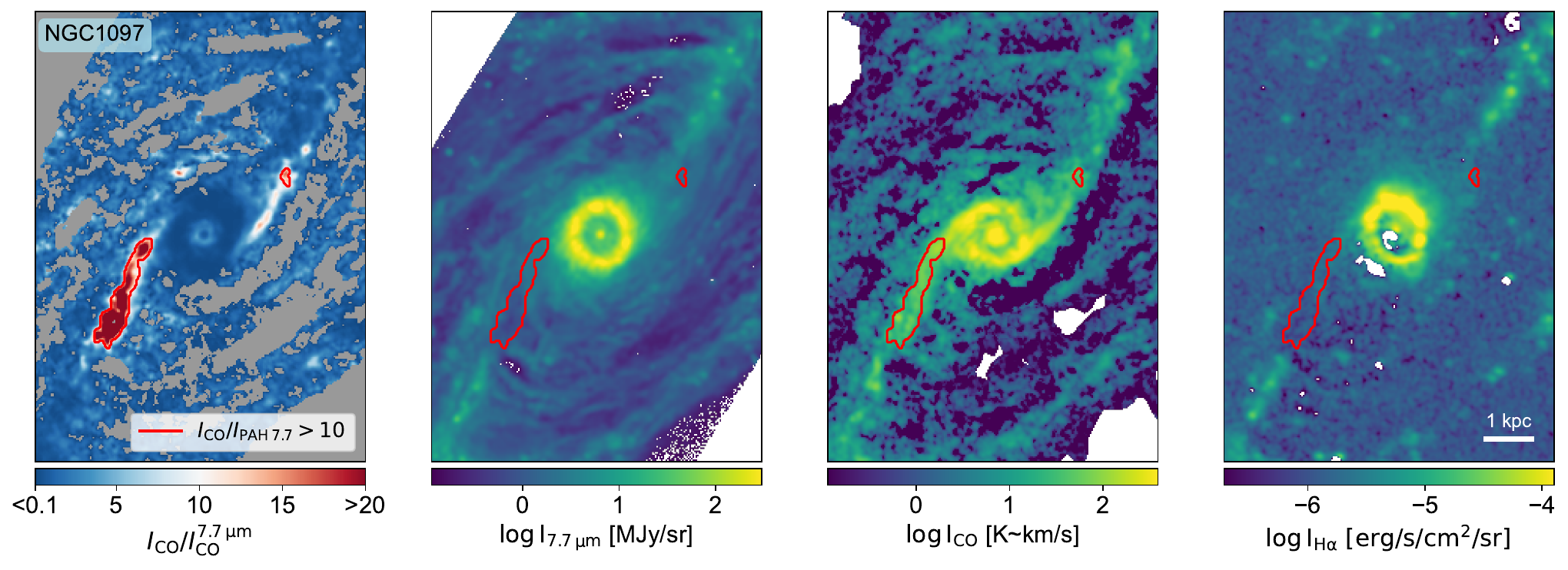}
\caption{Three examples of galaxies (NGC\,1566, NGC\,4303, and NGC\,1097) hosting outlier regions with abnormally high CO-to-PAH ratios. The first column shows the ratio of CO flux ($I_{\rm CO}$) to that expected from 7.7$\mu$m PAH emission ($I_{\mathrm{CO}}^{\mathrm{7.7\mu m}}$), using the relations of \citet{chown24}. We apply the same intensity cut as \citet{chown24}, including only pixels likely dominated by molecular gas with $I_{7.7\,\mu m}>0.5\mathrm{MJy\,sr^{-1}}$, and excluded regions are shown in gray. Outlier regions, defined as having a flux ratio ($I_{\mathrm{CO}}/I_{\mathrm{CO}}^{\mathrm{7.7\mu m}}$) higher than 10 (see text), are highlighted in red solid line. These outlier regions are spatially localized along bar lanes, spiral arms, and galaxy centers with no signs of massive star formation in H$\alpha$. Subsequent columns show the 7.7$\mu$m PAH, CO and H$\alpha$ observations. For NGC\,1566 and NGC\,4303, the H$\alpha$ maps are from PHANGS–MUSE data corrected for extinction \citep{belfiore22}, while for NGC\,1097 the H$\alpha$ map is from narrowband imaging that has not been corrected for extinction (Razza et al. in prep.).} \label{fig:fig1}
\end{figure*}

\subsection{Convolution and homogenization to a common pixel grid}
To proceed with our analysis of comparing CO and PAH emission, we apply additional processing to the data. First, the higher-resolution JWST observations are convolved to match the coarser resolution of the CO maps. The convolution uses a kernel that transforms the JWST PSF into a circular Gaussian PSF, generated using the code \texttt{jwst\_kernels}\footnote{\url{https://github.com/francbelf/jwst_kernels}}, following the method of \citet{aniano11} and \citet{williams24}. Then, the convolved images are re-gridded to align with the pixel grid of the CO data. 

\section{Outlier detection}\label{sec:result}

We perform outlier detection using CO and stellar continuum subtracted F770W maps that has been resampled on a gird with one CO-beam spacing to reduce inter-sample covariance. We then identify pixels whose CO emission is anomalously high relative to that expected from the 7.7\,$\mu$m PAH emission. For each galaxy, we predict CO emission from the stellar–continuum–subtracted F770W emission ($I_{\mathrm {7.7\mu m}}$) using the relation:
\begin{equation}
\log I_{\mathrm{CO}}^{\mathrm{7.7\mu m}} = m \left[ \log I_{\mathrm{7.7\mu m}} - x_{0} \right] + b ,
\end{equation}
where the coefficients $m$, $b$, and $x_0$ are individual galaxy based fits from Table~4 of \cite{chown24}. We then calculate residuals
\begin{equation}
\Delta = \log I_{\rm CO} - \log I_{\rm CO}^{7.7\mu m} = \log\left(\frac{I_{\rm CO}}{I_{\rm CO}^{7.7\mu m}}\right).
\end{equation}

Uncertainties on $\Delta$ ($\sigma_{\Delta}$) are obtained by error propagation, 
\begin{multline}
\sigma_{\Delta}^2=
\left(\frac{\sigma_{I_{\rm CO}}}{I_{\rm CO}\,\ln 10}\right)^{2}
+ \left(\frac{m\,\sigma_{I_{7.7\,\mu\mathrm{m}}}}{I_{7.7\,\mu\mathrm{m}}\,\ln 10}\right)^{2} \\
+ \big[(\log I_{\mathrm{7.7\mu m}} - x_0)\,\sigma_m\big]^2
+ \sigma_b^2
+ \sigma_{\rm int}^2 \, ,
\end{multline}
where $\sigma_{I_{\rm CO}}$ and $\sigma_{I_{7.7\,\mu m}}$ are per beam measurement errors, $(\sigma_m,\sigma_b)$ are the fit–parameter uncertainties, and $\sigma_{\rm int}$ is the intrinsic scatter of the relation from \citet{chown24}.

To remain consistent with \citet{chown24}, we limit the analysis to beam samples with $I_{\mathrm{7.7\mu m}}>0.5\,\mathrm{MJy/sr}$. To avoid noise contamination and isolate significant outliers, we additionally require $S/N>3$ in both $I_{\mathrm{CO}}$ and $I_{\mathrm{7.7\mu m}}$. We then identify beam samples with $I_{\mathrm{CO}}/I_{\mathrm{CO}}^{\mathrm{7.7\mu m}}>10$ and group adjacent beams into a contiguous region and compute inverse–variance–weighted statistics to calculate average residual and its uncertainty ($\bar{\Delta}$ and $\sigma_{\bar{\Delta}}$). We define a contiguous region as having abnormally high CO-to-PAH ratio if it satisfies the conservative detection threshold of $\bar{\Delta}/\sigma_{\bar{\Delta}}\ge10$. This results in a detection of outlier regions in 20 galaxies out of 70 galaxies analyzed in \citet{chown24}. If we relax the threshold to $\bar{\Delta}/\sigma_{\bar{\Delta}}\ge5$, additional outlier regions that are relatively small, composed of two to three CO beams, emerge in NGC\,1512, NGC\,1672, NGC\,2835, NGC\,4457, and NGC\,4548. Throughout the paper, we adopt a more conservative threshold $\bar{\Delta}/\sigma_{\bar{\Delta}}\ge10$ to focus on significant and obvious outliers.

As an additional robustness check, we varied the flux ratio criterion applied to beam samples from $I_{\mathrm{CO}}/I_{\mathrm{CO}}^{\mathrm{7.7\mu m}}>8$ to $>12$. Lowering the cut to 8 identifies additional outlier regions in NGC\,1512, NGC\,1672, NGC\,1792, NGC\,4457, NGC\,4548, and NGC\,7496. On the other hand, adopting a stricter cut of 12 removes smaller outliers and causes a few galaxies that previously hosted outliers to no longer meet the criterion (NGC\,3351, NGC\,4321, NGC\,4731, and NGC\,4941). Nonetheless, the key qualitative trends and our main conclusions remain unchanged.

Using the same procedure, we searched for low CO-to-PAH ratio outliers $I_{\mathrm{CO}}/I_{\mathrm{CO}}^{\mathrm{7.7\mu m}}<0.1$, which could signal CO-dark molecular or atomic gas \citep{sandstrom23,kim25}. We did not identify such regions, likely because the CO data is too shallow and noisy to establish a significant deficit relative to the PAH emission.

In Figure~\ref{fig:fig1}, we show the measured ratio ($I_{\mathrm{CO}}/I_{\mathrm{CO}}^{\mathrm{7.7\mu m}}$) for three galaxies, selected because their outlier regions are either spatially extended or located across diverse environments. The red contours in Figure~\ref{fig:fig1} highlight regions where the observed CO flux exceeds the value expected from the 7.7$\mu$m PAH emission by more than an order of magnitude. In Appendix~\ref{app:all}, we present the same figure for all 20 galaxies with such outliers, along with their properties in Table~\ref{tab:prop}, among the 70 galaxies included in our analysis. We note that most galaxies hosting outliers are classified as strongly or weakly barred (17 out of 20; \citealt{buta15}). However, outliers are not ubiquitous among barred systems, with only $\sim$35\% of barred galaxies exhibiting outliers.

These outlier regions represent significant deviations from the tight CO–7.7$\mu$m PAH relation reported by \citet{chown24}. They constitute 0.03-7\% of the total CO-emitting area with an average of $\sim$1\% (see Appendix~\ref{app:all}) and are spatially localized along bar lanes and centers without signs of massive star formation detected in H$\alpha$. NGC\,1566 is the only galaxy where we find such regions at the edges of spiral arms. The preferential location of these regions within specific dynamical structures suggests that they are not noise-driven statistical outliers. Instead, the elevated CO-to-PAH ratios have a physical origin linked to local gas dynamics and/or radiation environments.

In order to investigate the origin of these high CO-to-PAH ratio regions, we define a ``null-test'' control region with three requirements: 1) it lies near the outlier region within the same dynamical structure; 2) it contains only pixels with normal CO-to-PAH ratios ($I_{\mathrm{CO}}/I_{\mathrm{CO}}^{\mathrm{7.7\mu m}}<10$); and 3) it shows no obvious signs of recent star formation, so as to remain consistent with the outlier regions. Recent star formation is traced by extinction-corrected MUSE H$\alpha$ for Cycle~1 galaxies and by narrowband H$\alpha$ imaging for Cycle~2 galaxies. In addition to these, whenever possible, we also required the null-test control region to have a CO flux comparable to that of the outlier region. In practice, we select null-test regions near the outlier regions within the same dynamical structure, which closely matches the local environment and the galactocentric radius. This also allows us to mitigate the effect of metallicity gradients in our comparison between null-test and outlier region\footnote{Even if there is some metallicity differences between the null-test and outlier regions, both CO and PAH emission are expected to increase at higher metallicity due to a more efficient shielding and a higher PAH fraction \citep[e.g.,][]{engelbracht05,  chown25_lowz}, making metallicity variations an unlikely driver of the most extreme CO/PAH outliers.}. In some cases we do not choose the immediately neighboring area in order to avoid star-forming sites, but we still aim to select regions with comparable CO flux. The null-test regions therefore remain at similar galactocentric radii, show no obvious signs of recent star formation, and have sufficient CO signal for a meaningful comparison. The locations of these null test regions are shown in Appendix~\ref{app:all}.

In the following section, by comparing outlier regions to its null-test regions, we explore possible physical mechanisms, including suppressed PAH emission and enhanced CO emissivity that could produce such extreme deviations from the typical CO–PAH relation.

\section{Origin of outliers}\label{sec:discussion}

\begin{figure*}
\centering
\includegraphics[width=\textwidth]{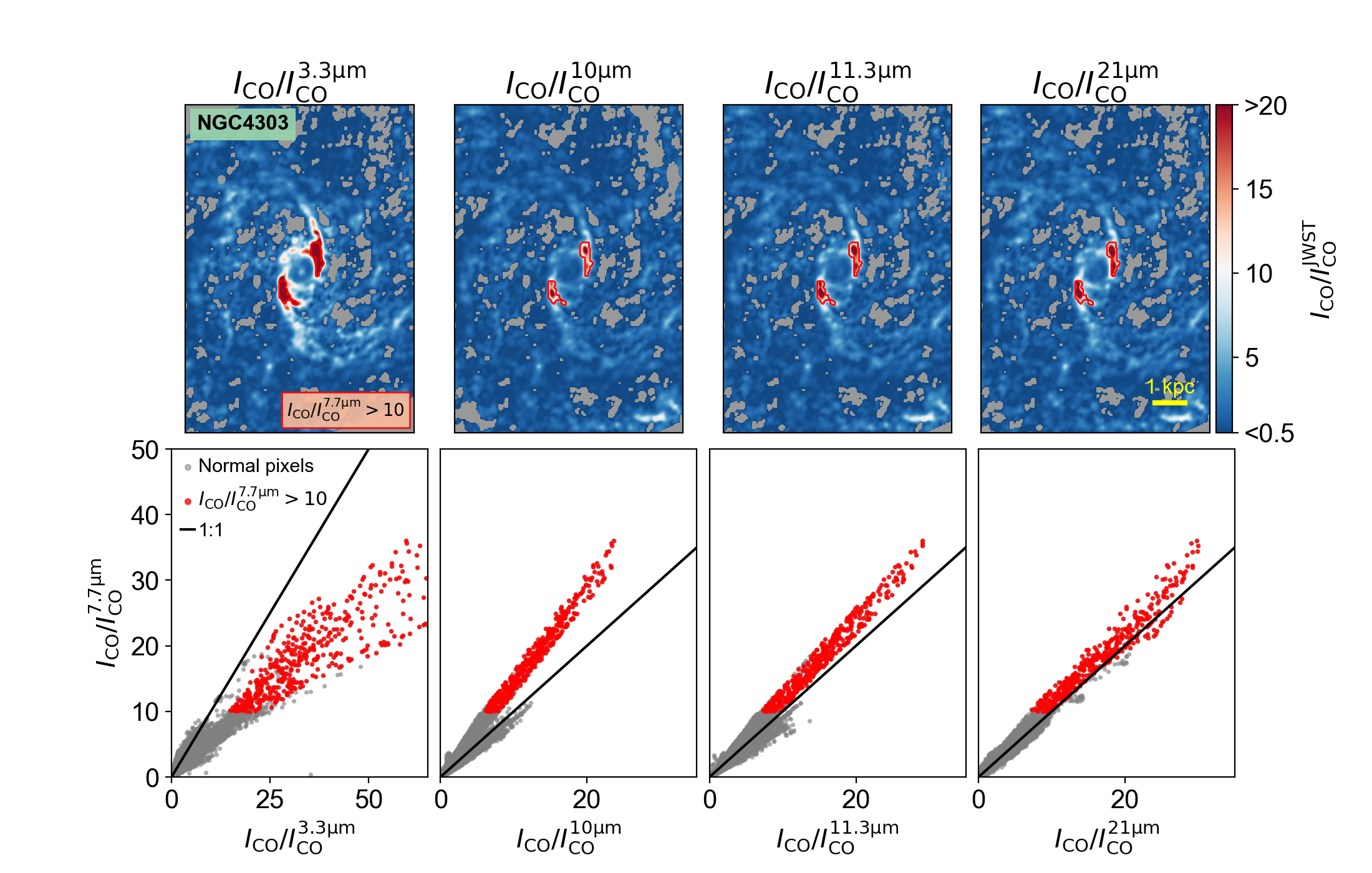}
\caption{The upper panels show the ratio of CO flux to that expected from other PAH- and dust continuum-dominated bands ($I_{\mathrm{CO}}/I_{\mathrm{CO}}^{\mathrm{JWST}}$), based on correlation coefficients reported in \citet[][for 3.3$\mu$m and 11.3$\mu$m]{chown24} and measured in this work (for 10$\mu$m and 21$\mu$m). Red contours highlight outlier regions identified using the 7.7$\mu$m band. The lower panel shows the relation between $I_{\mathrm{CO}}/I_{\mathrm{CO}}^{7.7\mu\mathrm{m}}$ and the corresponding ratios derived from other JWST bands. Data points within the outlier regions (defined by extreme $I_{\mathrm{CO}}/I_{\mathrm{CO}}^{7.7\mu\mathrm{m}}$ ratios) are shown in red, whereas all other pixels are shown in gray. \label{fig:otherbands}}
\end{figure*}

\subsection{Suppressed PAH Emission Scenario}\label{ssec:pahfaint}

As described in Section~\ref{ssec:data_jwst}, the JWST Cycle~1 data provide coverage across multiple filters, enabling measurements of key PAH emission features (at 3.3, 7.7, and 11.3$\mu$m) as well as the dust continuum (at 10$\mu$m and 21$\mu$m). Here, we investigate whether the outlier regions identified in the CO-to-7.7$\mu$m PAH ratio persist when other PAH or dust continuum dominated bands are used. We also compare PAH properties between the outlier and null-test regions.

\subsubsection{CO-to-mid-IR ratio using other PAH and dust continuum emission}
Figure~\ref{fig:otherbands} shows the CO flux relative to that expected from other PAH- and dust continuum–dominated bands in NGC\,4303, which is one of the galaxies with the most prominent CO-to-7.7$\mu$m PAH anomalies. We estimate CO fluxes from 3.3 and 11.3$\mu$m bands using the coefficients from \citet{chown24} ($I_{\mathrm{CO}}/I_{\mathrm{CO}}^{3.3\mu m}$ and $I_{\mathrm{CO}}/I_{\mathrm{CO}}^{11.3\mu m}$), and from 10 and 21$\mu$m bands using coefficients derived in this work following the same method ($I_{\mathrm{CO}}/I_{\mathrm{CO}}^{10\mu m}$ and $I_{\mathrm{CO}}/I_{\mathrm{CO}}^{21\mu m}$).

The outlier regions identified in CO-to-7.7$\mu$m PAH ratio (red) consistently show elevated ratios across all PAH and continuum bands, with the largest excess observed at 3.3$\mu$m and the smallest at 10$\mu$m. Other galaxies hosting outlier regions also show a similar behavior. The significant enhancement across multiple bands indicates that 7.7$\mu$m is not particularly suppressed and the anomaly is more likely due to overluminous CO emission rather than a physical mechanism that suppresses multiple PAH bands and dust continuum emission at the same time. It is also interesting to note that $I_{\mathrm{CO}}/I_{\mathrm{CO}}^{\mathrm{7.7\mu m}}$ shows the closest agreement with $I_{\mathrm{CO}}/I_{\mathrm{CO}}^{\mathrm{21\mu m}}$, even though $7.7\mu$m traces mostly PAH emission while $21\mu$m traces mostly dust continuum. Further supporting this picture, Appendix~\ref{app:all} shows three–color HST composites (F814W, F438W, F555W; \citealp{lee22}) in which the outlier regions display dust extinction similar to their surroundings (see also \citealt{faustino-viera25}). This suggests that outlier regions exhibit comparable dust content compared to their surroundings, making simultaneous suppression of PAH and dust emission, for example due to shielding, an unlikely explanation for the elevated CO/PAH ratios.

\begin{figure*}
\centering
\includegraphics[scale=0.7]{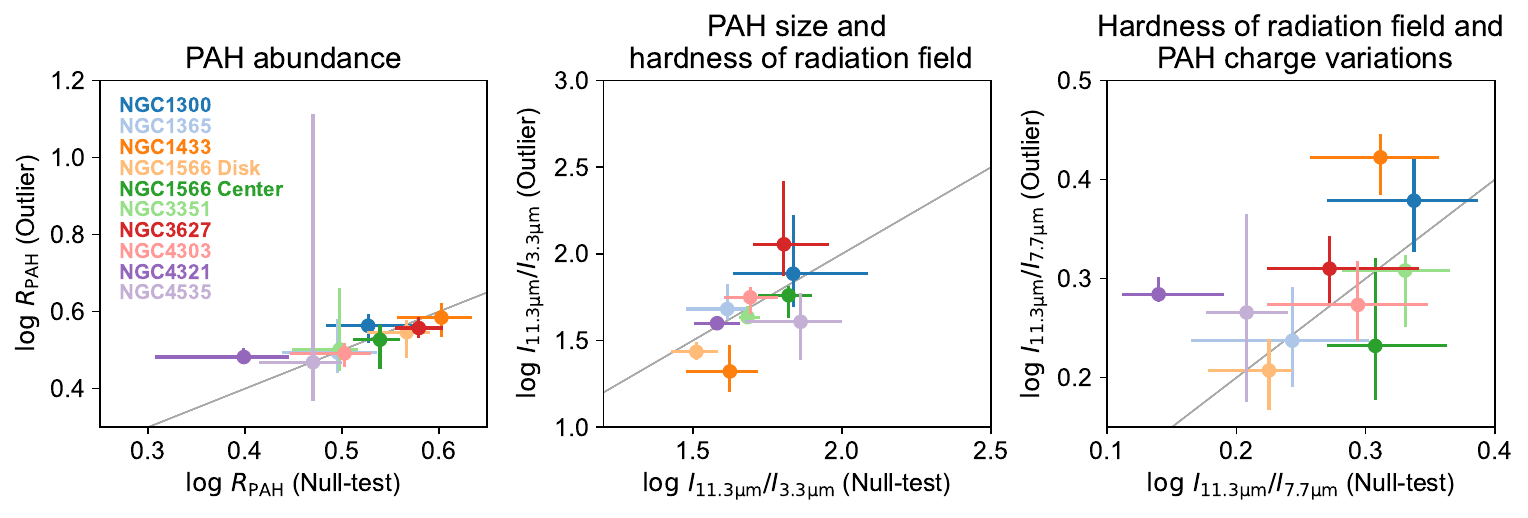}
\caption{Comparison of PAH properties between outlier and null-test regions in the JWST Cycle~1 sample. The first panel shows the median and 1$\sigma$ distribution of $R_{\mathrm{PAH}}$ measured in outlier and null-test regions. The second and third panels show $I_{\mathrm{11.3\mu m}}/I_{\mathrm{3.3\mu m}}$, which is sensitivity to the PAH size distribution and the hardness of the radiation field, and $I_{\mathrm{11.3\mu m}}/I_{\mathrm{7.7\mu m}}$, which trace the radiation field hardness and PAH charge states. The gray solid line indicates the one-to-one relation. As outlier regions exist both in center and disk in NGC\,1566, two environments are shown separately. \label{fig:pahprops}}
\end{figure*}

\subsubsection{Difference in PAH properties}

PAH intensity ratios are sensitive to molecule properties (e.g., size, asymmetry, composition, and charge state) and to local radiation field \citep{hony01, galliano08, lai20, Maragkoudakis20}. In particular, 11.3$\mu$m/3.3$\mu$m becomes lower as the spectral energy distribution of the illumination radiation becomes harder and PAH size distributions become smaller. The ratio of 11.3$\mu$m/7.7$\mu$m also decreases with harder radiation field and is very sensitive to changes of PAH charge variations \citep{lai22, chastenet23, draine21, hensley23_model, baron24, baron25}. Furthermore, the ratio of emission in PAH bands to the 21\,$\mu$m dust continuum is considered as a strong tracer of the fraction of dust mass in the form of PAHs \citep{chastenet19, sutter24, egorov23}.

In Figure~\ref{fig:pahprops}, we show median and 1$\sigma$ distribution of PAH band ratios to assess whether PAH properties differ between our outlier and null-test regions across Cycle~1 galaxies. The fraction of dust mass in PAHs, $R_{\mathrm{PAH}}$, estimated from $(I_{\mathrm{7.7\mu m}}+I_{\mathrm{11.3\mu m}})/I_{\mathrm{21\mu m}}$, shows no significant difference between outlier and null-test regions. We note that the outlier and null-test regions also do not overlap with massive star-forming sites, where $R_{\mathrm{PAH}}$ is typically suppressed due to PAH destruction by stellar feedback \citep{chastenet23, egorov23, Egorov2025, sutter24}.

In the middle panel, we compare $I_{\mathrm{11.3\mu m}}/I_{\mathrm{3.3\mu m}}$. While most galaxies follow a one-to-one relation, outlier regions in NGC\,1433 and NGC\,4535 show somewhat lower $I_{\mathrm{11.3\mu m}}/I_{\mathrm{3.3\mu m}}$ compared to their null-test region, suggesting fragmentation into smaller grains or harder radiation field. In contrast, NGC\,3627, shows a higher ratio in its outlier region (see also \citep{dale25}), possibly indicating PAH growth or coagulation onto dust grains, making 7.7$\mu$m and 11.3$\mu$m bands fainter \citep{draine21, hensley23_model}. As most of these outlier regions lie along bar lanes with no signs of massive star formation, the differences in $I_{\mathrm{11.3\mu m}}/I_{\mathrm{3.3\mu m}}$ suggest that PAH size distributions may be altered by shocks or coagulation as gas and dust flow inward. 

Lastly, we compare $I_{\mathrm{11.3\mu m}}/I_{\mathrm{7.7\mu m}}$ PAH ratios. The $I_{\mathrm{11.3\mu m}}/I_{\mathrm{7.7\mu m}}$, often used as a tracer of PAH ionization \citep[e.g.,][]{smith07, lai22}, has also been shown to correlate with $\mathrm{[SII]/H\alpha}$ \citep{baron24}. This implies that the majority of the variation in $I_{\mathrm{11.3\mu m}}/I_{\mathrm{7.7\mu m}}$ ratios is driven by the changing shape of the illuminating spectral energy distribution, where the ratio becomes higher as the radiation from old stellar population dominates (high $I_{\mathrm{11.3\mu m}}/I_{\mathrm{7.7\mu m}}$ and high $\mathrm{[SII]/H\alpha}$) compared to the radiation from young massive stars (low $I_{\mathrm{11.3\mu m}}/I_{\mathrm{7.7\mu m}}$ and low $\mathrm{[SII]/H\alpha}$). In NGC\,1433 and NGC\,4321, we find somewhat higher $I_{\mathrm{11.3\mu m}}/I_{\mathrm{7.7\mu m}}$ in outlier regions ($\sim0.1$\,dex), which may indicate an older stellar population dominating the radiation and/or more neutral PAHs, and therefore the lack of PAH emission compared to CO. However, in most galaxies, outlier and null-test regions agree within 1$\sigma$.

While some outlier regions show hints of altered PAH sizes or hardness of the radiation field, these trends are not consistent across systems. As the PAH ratios explored here are degenerate with multiple physical effects, additional spectroscopic tracers (e.g., H$_2$ rotational and rovibrational lines, [Fe II] for shocks, and a wider suite of PAH features) will be essential to further investigate whether PAH emission is suppressed in these regions.

\subsection{Elevated CO emissivity scenario}\label{ssec:highco}

Almost all outlier regions lie along the dust lanes in bars, sites where inflowing gas experiences streaming motions and strong shocks due to intersecting orbital families, causing abrupt changes in velocity \citep{roberts79, athanassoula92, sormani15, sellwood93}. These dynamical processes can elevate the CO velocity dispersion and increase the integrated intensity either through line broadening or by reducing the optical depth, which boosts the CO emissivity \citep{bolatto13, gong20, teng23}. In addition, shocks and strong shear can raise the excitation temperature and collision rates, populating higher-$J$ levels and further enhancing the CO flux for a given molecular gas mass.

In this Section, we test whether the high CO-to-PAH ratios arise from elevated CO emissivity or enhanced excitation by comparing CO velocity dispersions and CO line ratios between the outlier and null-test regions. If the high CO/PAH ratios are associated with higher velocity dispersions or elevated excitation temperature, it would indicate that the $\Sigma_{\rm H_{2}}$-to-PAH ratio between outlier and null-test regions should remain similar and the CO–to–$\mathrm{H}_2$ conversion factor ($\alpha_{\rm CO}$) should be at least 10 times lower in the outlier regions compared to the regions with $I_{\rm CO}/I_{\rm CO}^{7.7\,\mu\rm m}=1$.
 
\begin{figure*}
\centering
\includegraphics[width=\textwidth]{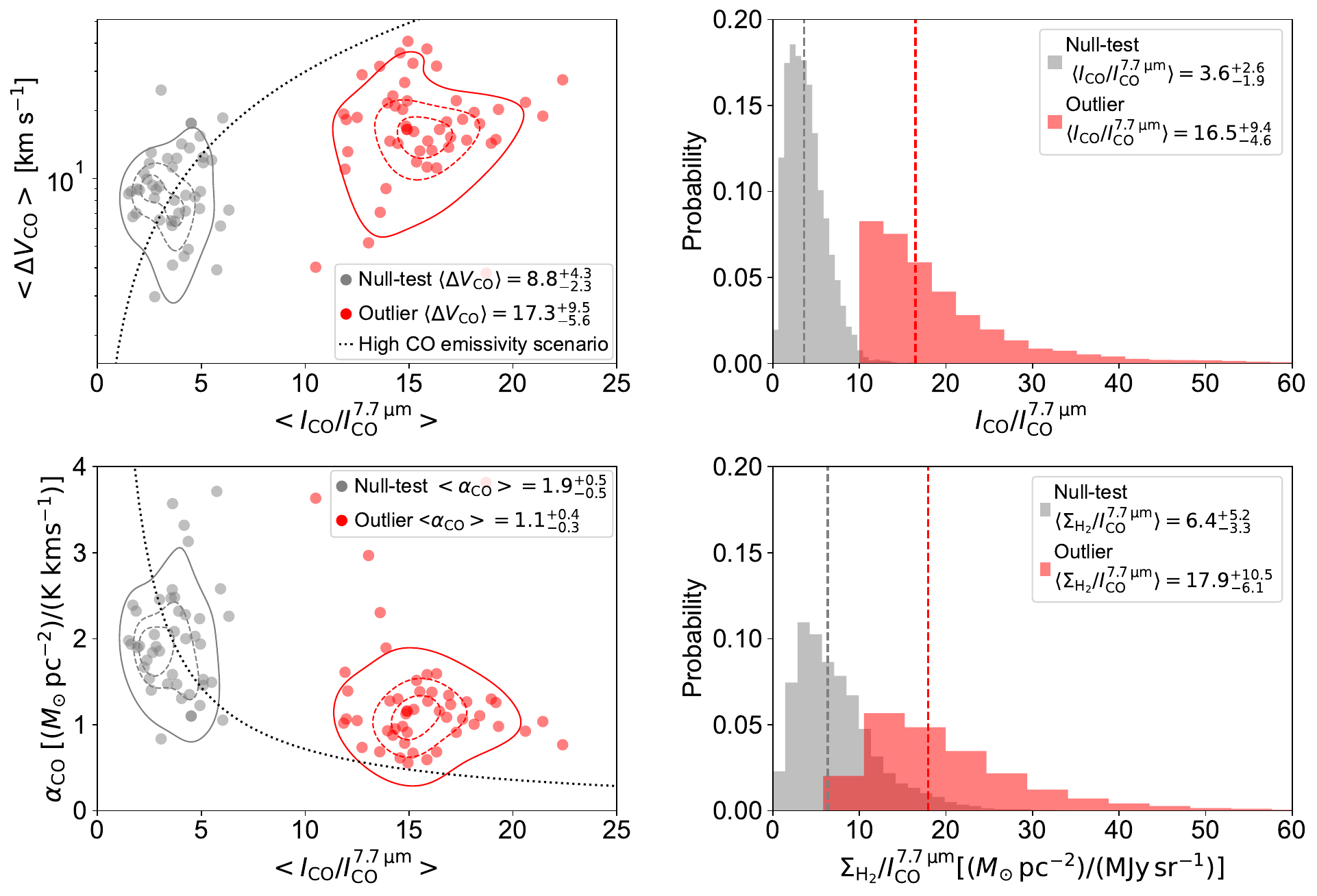}
\includegraphics[width=\textwidth]{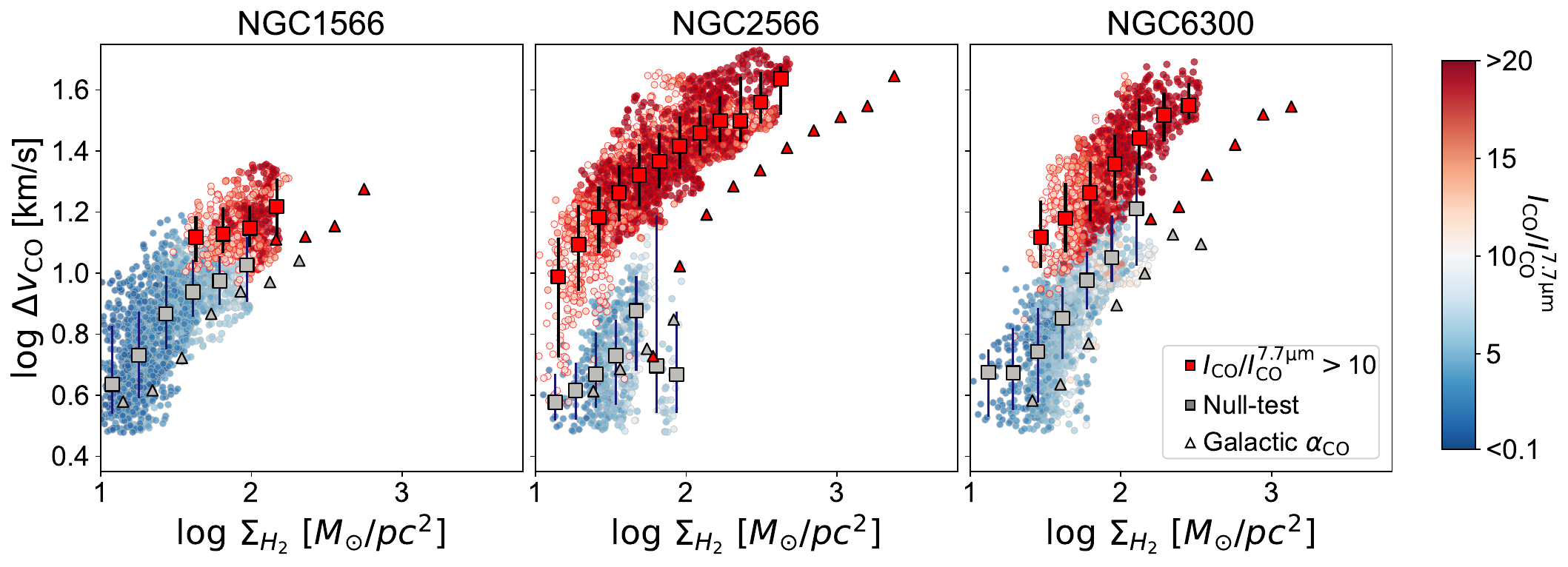}
\caption{\textit{Top left:} CO luminosity-weighted velocity dispersion ($\langle\Delta v_{\rm CO}\rangle$) is higher in outlier regions compared to their null-test regions. $\langle\Delta v_{\rm CO}\rangle$ versus the median CO-to-PAH ratio, $\langle I_{\rm CO}/I_{\rm CO}^{7.7\,\mu{\rm m}}\rangle$, for each outlier (red) and null-test (gray) region across the 20 galaxies hosting CO/PAH outliers. The outermost contour encloses 68\% of the points (solid line; 1$\sigma$). The dotted black line shows the $\langle\Delta v_{\rm CO}\rangle$ required to explain the outliers solely via enhanced CO emissivity at fixed H$_2$/PAH. \textit{Top right:} Distributions of $I_{\rm CO}/I_{\rm CO}^{7.7,\mu{\rm m}}$ for outlier (red) and null-test (gray) regions. \textit{Middle left:} As in the upper panel, $\alpha_{\rm CO}$ (from the \citealt{teng24} prescription) is shown as a function of $\langle I_{\rm CO}/I_{\rm CO}^{7.7,\mu{\rm m}}\rangle$. \textit{Middle right:} Distributions of $\Sigma_{\rm H_2}/I_{\rm CO}^{7.7,\mu{\rm m}}$, where $\Sigma_{\rm H_2}=I_{\rm CO}\,\alpha_{\rm CO}$. \textit{Bottom:} $\Delta v_{\mathrm{CO}}$ as a function of molecular gas surface density ($\Sigma_{\mathrm{H_{2}}}$) for  NGC\,1566, NGC\,2566, and NGC\,6300. Data points are color-coded by the observed-to-expected CO flux ratio ($I_{\mathrm{CO}}/I_{\mathrm{CO}}^{7.7,\mu{\rm m}}$), where points associated with outliers are outlined in red. We show the median $\Delta v_{\mathrm{CO}}$ in logarithmically spaced bins of $\log \Sigma_{\mathrm{H_{2}}}$ together with 1$\sigma$ scatter for each outlier and null test region (red and gray squares, respectively). While $\Sigma_{\mathrm{H_{2}}}$ is estimated using the $\alpha_{\mathrm{CO}}$–$\Delta v_{\mathrm{CO}}$ relation from \citet{teng24}, we also show the median of $\Delta v_{\mathrm{CO}}$ as a function of $\Sigma_{\mathrm{H_{2}}}$ when a constant galactic $\alpha_{\mathrm{CO}}=4.35 \mathrm{M_{\odot}pc^{-2}/(K~km~s^{-1})}$ is adopted instead (triangles). }
\label{fig:vdisp_all}
\end{figure*}

\subsubsection{CO velocity dispersion}\label{sssec:vdisp}
Centers of barred galaxies exhibit higher velocity dispersions compared to the galaxy disk or centers of non-barred galaxies \citep{sun20}. The elevated dispersions in barred environments reduce optical depth, raise the CO emissivity, and lower the $\alpha_{\mathrm{CO}}$ \citep{teng24}. Using dust-based, spatially resolved $\alpha_{\mathrm{CO}}$ measurements at $\sim$2\,kpc scale for nearby PHANGS galaxies \citep{chiang24}, \citet{teng24} found a strong anti-correlation between $\alpha_{\mathrm{CO}}$ and velocity dispersion across both galaxy centers and disks, suggesting that the CO velocity dispersion directly traces changes in CO emissivity. 

Figure~\ref{fig:vdisp_all} shows the CO luminosity-weighted average velocity dispersion ($\langle \Delta v_{\mathrm{CO}} \rangle$) as a function of the median CO-to-PAH ratio ($\langle I_{\mathrm{CO}}/I_{\mathrm{CO}}^{\mathrm{7.7\mu m}}\rangle$) of each outlier region and null-test region, across the 20 galaxies hosting CO/PAH outliers. On average, the outlier regions show two times higher $\Delta v_{\mathrm{CO}}$ and $\sim 5$ times higher $I_{\mathrm{CO}}/I_{\mathrm{CO}}^{\mathrm{7.7\mu m}}$ compared to those in null-test regions\footnote{When all pixels are considered and not just the null-test region, the distribution of $I_{\mathrm{CO}}/I_{\mathrm{CO}}^{\mathrm{7.7\mu m}}$ strongly peaks at 1.}. The bottom panel shows $\alpha_{\mathrm{CO}}$ as a function of the median CO-to-PAH ratio ($\langle I_{\mathrm{CO}}/I_{\mathrm{CO}}^{\mathrm{7.7\mu m}}\rangle$). We obtain $\alpha_{\mathrm{CO}}$ using the \citet{teng24} prescription,
\begin{equation}
    \log \alpha_{\mathrm{CO}}=-0.81 \log \langle\Delta v_{\mathrm{CO}} \rangle+1.05,
\end{equation}
with $ \langle\Delta v_{\mathrm{CO}}\rangle$ measured within each region. We also compare the distributions of $I_{\mathrm{CO}}/I_{\mathrm{CO}}^{\mathrm{7.7\mu m}}$ and $\Sigma_{\rm{H_{2}}}/I_{\mathrm{CO}}^{\mathrm{7.7\mu m}}$, where $\Sigma_{\rm{H_{2}}}=I_{\mathrm{CO}}\,\alpha_{\rm {CO}}$. As outlier regions are associated with higher velocity dispersions and thus lower $\alpha_{\mathrm{CO}}$, the large offset observed between the outlier and null-test regions in $I_{\mathrm{CO}}/I_{\mathrm{CO}}^{\mathrm{7.7\mu m}}$ distribution decreases when $\Sigma_{\rm{H_{2}}}=I_{\mathrm{CO}}\,\alpha_{\rm {CO}}$ is used. We also show how high $\Delta v_{\mathrm{CO}}$ (or how low the $\alpha_{\mathrm{CO}}$) should be in order to explain these outliers solely by enhanced CO-emissivity, under the assumption of a constant $\rm H_{2}$-to-PAH ratio (dotted line) between the outlier and null-test regions. While the outlier regions are associated with higher velocity dispersion, it appears it should be even higher (three times more) to explain these outliers from enhanced emissivity.

The bottom panel of Figure~\ref{fig:vdisp_all} shows CO velocity dispersion ($\Delta v_{\rm CO}$) as a function of molecular gas surface density ($\Sigma_{\rm H_2}$) for three galaxies where the enhancement in CO velocity dispersion is prominent. We also divide the data into logarithmically spaced bins in $\log~\Sigma_{\mathrm{H_{2}}}$ and then compute median and 1$\sigma$ scatter of $\Delta v_{\mathrm{CO}}$ in each bin for pixels in outlier null-test regions. Within the range of $\Sigma_{\mathrm{H_{2}}}$ where the outlier and null-test regions overlap, the outlier regions generally exhibit higher $\Delta v_{\mathrm{CO}}$, indicating higher CO emissivity that can lead to abnormally high CO-to-PAH ratios. We note that the measured $\Delta v_{\mathrm{CO}}$ can be inflated by unresolved velocity gradients in the galaxy centers where large-scale rotational shear is steep. However, as we compare outlier regions to nearby null-test regions within the same dynamical structure at similar galactocentric radii, our analysis is expected to be insensitive to beam smearing driven by the large-scale rotation. Moreover, any remaining broadening associated with non-circular motions (e.g., bar-driven streaming, shocks, and shear) is physically relevant to our interpretation.

In Appendix~\ref{app:vdisp_c2}, the same figure for all 20 galaxies hosting outlier regions is shown. While in some galaxies the velocity dispersions at a given $\Sigma_{\mathrm{H_{2}}}$ agree within $1\sigma$, almost half of them show differences exceeding this level, particularly NGC\,1566, NGC\,3351, NGC\,4321, NGC\,2566, NGC\,2997, NGC\,4569, NGC\,4579, and NGC\,6300. We also note that when $\alpha_{\rm CO}$ prescription from \citep{teng24} is used, $\Sigma_{\mathrm{H_{2}}}$ between outlier and null-test regions overlaps much more significantly than when a constant Galactic $\alpha_{\mathrm{CO}}=4.35\,\mathrm{M_{\odot}pc^{-2}/(K~km~s^{-1})}$ is adopted (triangles). This again supports the view where high CO-to-PAH ratios are driven by elevated CO emissivity. The outlier and null-test regions contain comparable amounts of PAH emission for a given amount of $\mathrm{H_{2}}$. These results would imply that the $\alpha_{\rm CO}$ is more than 10 times lower in the outlier region compared to the typical regions with a normal ratio of $I_{\rm CO}/I_{\rm CO}^{7.7\,\mu\rm m}=1$, suggesting a strong sub-kpc variation in $\alpha_{\rm CO}$ as observed in the Milky Way and centers of barred galaxies \citep{blitz1985, teng22, teng23}.

\subsubsection{CO excitation from multi-transition lines and CO isotopologues} \label{sssec:co_exc}

High-resolution observations of CO multi-transitions and CO isotopologues observations would enable direct constraints on variations in CO line opacity and emissivity. For example the ratio between CO lines from $J{=}2{-}1$ and $J{=}1{-}0$ ($R_{21}$) increases in more highly excited gas, where higher kinetic temperature and density populate $J{=}2$ more efficiently than $J{=}1$, enhancing CO($J{=}2{-}1$) relative to CO($J{=}1{-}0$). However these lines are optically thick in molecular clouds and the ratio saturates. CO isotopologues, which are less opaque than $^{12}$CO, further help disentangle excitation from opacity effects. Their multi-transition line ratios provide sensitive probes of excitation, while $^{12}$CO/$^{13}$CO ratios are sensitive to isotopologue abundance and/or optical depth. 

Using CO isotopologues observations in their low rotational transitions, \citet{teng22} and \citet{teng23} have measured local $\alpha_{\mathrm{CO}}$ variations in the barred-centers of NGC\,3351, NGC\,3627, and NGC\,4321, which were found to be 4-15 times lower compared to the Galactic value. These three galaxies overlap with our sample that host outlier regions, however, the current available data are not deep enough to detect $^{13}$CO in the regions of interest for NGC\,3627 and NGC\,4321. As for NGC\,3351, the locations of high CO/PAH ratio is near the region with very low $\alpha_{\mathrm{CO}}$, located at where bar-driven inflow feeds the nucleus.

\begin{figure*}
\centering
\includegraphics[width=
\textwidth]{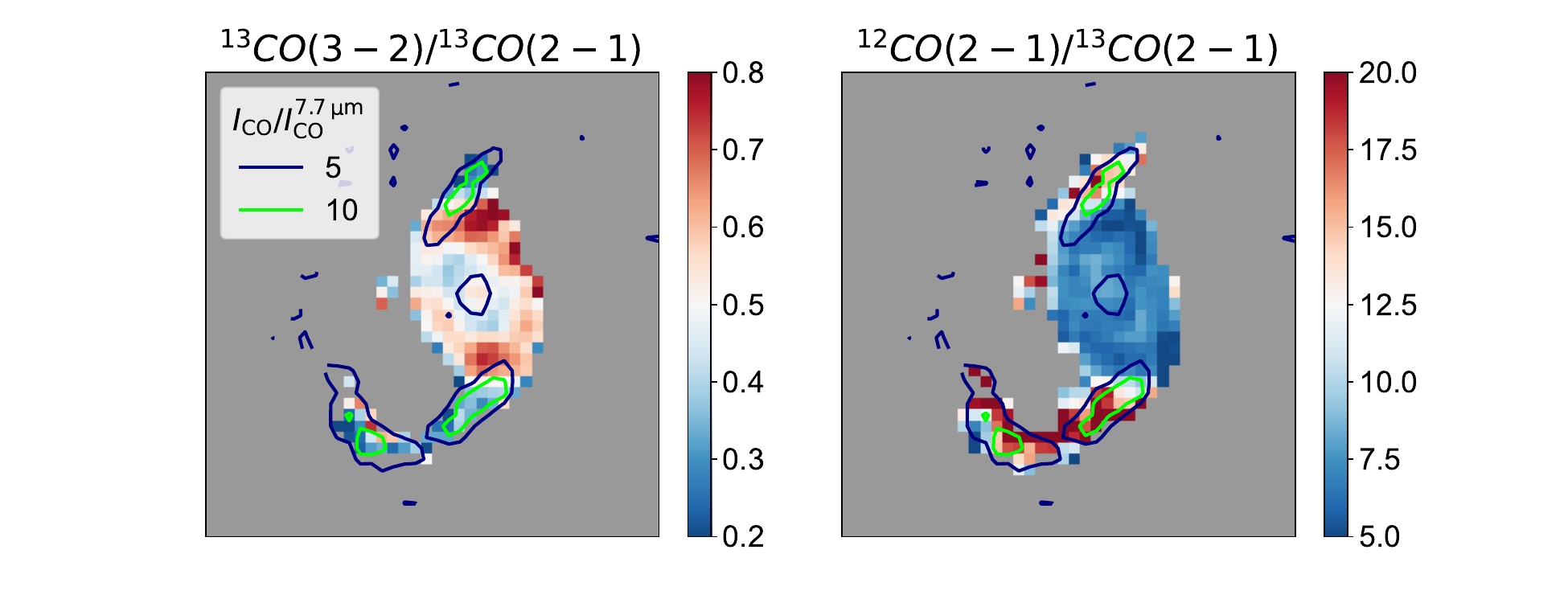}
\caption{CO line ratio maps of NGC\,3351 center from \citet{teng22}, overlaid with contour levels of $I_{\mathrm{CO}}/I_{\mathrm{PAH}}^{7.7\,\mu\mathrm{m}}$ at 5 and 10. \textit{Left:} $^{13}$CO($J{=}3{-}2$)/($J{=}2{-}1$), primarily sensitive to excitation temperature. \textit{Right:} $^{12}$CO/$^{13}$CO($J{=}2{-}1$), primarily sensitive to isotopologue abundance and/or optical depth. 
Regions with enhanced $I_{\mathrm{CO}}/I_{\mathrm{PAH}}^{7.7\,\mu\mathrm{m}}$ coincide with lower $^{13}$CO($3{-}2$)/($2{-}1$) and higher $^{12}$CO/$^{13}$CO($2{-}1$), indicating that the elevated CO-to-PAH ratio is not driven by higher excitation temperature. Instead, it is consistent with reduced optical depth, which allows more $^{12}$CO emission to escape and increases $^{12}$CO/$^{13}$CO($2{-}1$).
 \label{fig:t22ngc3351}}
\end{figure*}

Using $R_{21}$ maps from \citet[][ J. den Brok et al. in preparation]{denbrok25} for four galaxies hosting outliers (NGC\,1300, NGC\,1433, NGC\,2566, and NGC\,3627), we did not find significant differences in $R_{21}$ between our outlier and null-test regions exceeding 1$\sigma$. \citet{teng23} have mapped $^{13}$CO($J{=}3{-}2$)/($J{=}2{-}1$) $^{12}$CO/$^{13}$CO($J{=}2{-}1$) in NGC\,3351 including the outlier region. Figure~\ref{fig:t22ngc3351} shows that outlier regions are associated with lower $^{13}$CO($J{=}3{-}2$)/($J{=}2{-}1$) and higher $^{12}$CO/$^{13}$CO($J{=}2{-}1$). These trends suggest that the high CO-to-PAH ratio is not due to higher excitation temperature and is consistent with reduced optical depth that allows more CO emission to escape, thereby boosting both the integrated CO intensity and the $^{12}$CO/$^{13}$CO($J{=}2{-}1$) ratio.

\section{Summary}\label{sec:summary}
Under the conditions that dust and PAH grains are well mixed with gas and variations in the strength of the interstellar radiation field are modest, stochastically-heated PAH emission is expected to show correlation with the total amount of gas, often traced with CO in molecular gas-dominated galaxies. In line with this expectation, \citet{chown24} have found a tight and almost linear correlation between 7.7$\mu$m PAH emission and CO flux on 100\,pc scales, across 70 PHANGS-JWST galaxies. 

In this work, we present localized departures from this CO–7.7$\mu$m PAH relation, highlighting regions where the tight correlation breaks down. We have identified regions where the observed CO flux is more than 10 times brighter than that expected from 7.7$\mu$m PAH emission. We find that such outliers are common, observed in 20 galaxies out of 70 explored in \citet{chown24}. These regions with abnormally high $I_{\mathrm{CO}}/I_{\mathrm{CO}}^{\mathrm{7.7\mu m}}$ are mostly located along bar lanes and in the centers of barred galaxies with an exception of NGC\,1566 where we find such regions to exist in the concave part of the spiral arm as well. Interestingly, outlier regions do not show associated H$\alpha$ emission, indicating that the high CO-to-PAH ratio is not due to massive stars dissociating PAHs \citep{egorov23, sutter24}. 

We investigate the origin of the abnormally high CO-to-PAH ratios, which may arise from either suppressed 7.7$\mu$m PAH emission or enhanced CO emissivity.\\

\noindent\textbf{1. Suppressed 7.7$\mu$m PAH emission scenario:}
\begin{itemize} 
    \item For the JWST Cycle~1 galaxies, we find that the outlier regions, identified based on elevated CO-to-7.7$\mu$m PAH ratios, also consistently exhibit elevated ratios when using other PAH (3.3 and 11.3$\mu$m) and dust continuum bands (10 and 21$\mu$m). The fact that the ratios are significantly enhanced across multiple PAHs and dust-continuum bands suggests that the high CO-to-PAH ratios are more likely due to overluminous CO emission, rather than suppressed 7.7$\mu$m PAH emission.

    \item We also find that some outlier regions either show bigger or smaller PAH size distributions or hints of PAHs being excited by softer radiation fields compared to their null-test regions. However, these trends are not consistent across the system. 
\end{itemize}

\noindent\textbf{2. High CO emissivity scenario:}
\begin{itemize}

    \item Outlier regions generally show higher CO velocity dispersions ($\Delta v_{\mathrm{CO}}$) compared to their null-test regions. Higher velocity dispersion leads to lower optical depth in the CO line, thereby increasing the CO emissivity for a given gas mass. This implies that the outlier regions contains a similar amount of gas as their neighboring regions (as suggested by the similar PAH flux), but the elevated CO emissivity enhances the CO-to-PAH ratio. 

    \item Multi-transition CO and isotopologue data enables more direct constraints on opacity and excitation. While $R_{21}$ can rise with higher excitation, $R_{21}$ maps for four galaxies show no differences exceeding 1$\sigma$ between outlier and null-test regions. In NGC\,3351, isotopologue ratios show lower $^{13}\mathrm{CO}(3{-}2)/(2{-}1)$ and higher $^{12}\mathrm{CO}/^{13}\mathrm{CO}(2{-}1)$ in outlier regions, favoring reduced effective optical depth as the driver of the high CO/PAH instead of higher excitation temperature.

    \item These results support the idea that variations in $\alpha_{\mathrm{CO}}$ can occur on sub-kpc scales with changes of up to an order of magnitude, as observed in the Milky Way \citep{blitz1985} and centers of barred galaxies \citep{teng22, teng23}, and $\alpha_{\mathrm{CO}}$ is not solely determined by large-scale galactic environments (e.g., bar versus disk) or metallicity. 
    
\end{itemize} 
    
In summary, we have explored two possible scenarios to explain the extreme outliers in CO-PAH relation. While the elevated CO velocity dispersion may suggest changes in CO emissivity as the driver of the enhanced CO-to-PAH ratios, this interpretation requires confirmation through deep observations of CO isotopologues and multi-transitions that can directly constrain variations in optical depth and excitation. Future JWST IFU observations targeting additional PAH bands, [FeII] shock tracers, Hydrogen recombination lines, and H$\mathrm{_2}$ rotational and rovibrational lines, will help constrain the physical conditions of the PAHs, identify shocks which can destroy PAHs, and map the warm molecular gas, thereby distinguishing whether the CO emission is enhanced by the local environment.

\begin{acknowledgments}
We thank the referee for helpful comments, which improved the quality of the manuscript. This work has been carried out as part of the PHANGS collaboration. This work is based on observations made with the NASA/ESA/CSA JWST. The data were obtained from the Mikulski Archive for Space Telescopes at the Space Telescope Science Institute, which is operated by the Association of Universities for Research in Astronomy, Inc., under NASA contract NAS 5-03127 for JWST. These observations are associated with programs 2107 and 3707. 

J.K. is supported by a Kavli Fellowship at the Kavli Institute for Particle Astrophysics and Cosmology (KIPAC). S.E.C. acknowledges support from the National Science Foundation under grant No. AST-2441452, and from an Alfred P. Sloan Research Fellowship.
K.K. acknowledges support from the Deutsche Forschungsgemeinschaft (DFG, German Research Foundation) in the form of an Emmy Noether Research Group (grant number KR4598/2-1, PI Kreckel) and the European Research Council’s starting grant ERC StG-101077573 (“ISM-METALS"). OE acknowledges funding from the Deutsche Forschungsgemeinschaft (DFG, German Research Foundation) -- project-ID 541068876.
S.K.S is supported by a International Research Fellowship of Japan Society for the Promotion of Science (JSPS).

This paper makes use of the following ALMA data, which have been processed as part of the PHANGS--ALMA survey: \\
\noindent ADS/JAO.ALMA\#2012.1.00650.S, \linebreak 
ADS/JAO.ALMA\#2013.1.00803.S, \linebreak 
ADS/JAO.ALMA\#2013.1.01161.S, \linebreak 
ADS/JAO.ALMA\#2015.1.00121.S, \linebreak 
ADS/JAO.ALMA\#2015.1.00782.S, \linebreak 
ADS/JAO.ALMA\#2015.1.00925.S, \linebreak 
ADS/JAO.ALMA\#2015.1.00956.S, \linebreak 
ADS/JAO.ALMA\#2016.1.00386.S, \linebreak 
ADS/JAO.ALMA\#2017.1.00392.S, \linebreak 
ADS/JAO.ALMA\#2017.1.00766.S, \linebreak 
ADS/JAO.ALMA\#2017.1.00886.L, \linebreak 
ADS/JAO.ALMA\#2018.1.01321.S, \linebreak 
ADS/JAO.ALMA\#2018.1.01651.S, \linebreak 
ADS/JAO.ALMA\#2018.A.00062.S, \linebreak 
ADS/JAO.ALMA\#2019.1.01235.S, \linebreak 
ADS/JAO.ALMA\#2019.2.00129.S, \linebreak 
ALMA is a partnership of ESO (representing its member states), NSF (USA), and NINS (Japan), together with NRC (Canada), NSC and ASIAA (Taiwan), and KASI (Republic of Korea), in cooperation with the Republic of Chile. The Joint ALMA Observatory is operated by ESO, AUI/NRAO, and NAOJ. The National Radio Astronomy Observatory is a facility of the National Science Foundation operated under cooperative agreement by Associated Universities, Inc.
\end{acknowledgments}

%
\facilities{JWST, VLT/MUSE, ALMA}

\software{Astropy \citep{astropy:2013, astropy:2018, astropy:2022}, Matplotlib \citep{Hunter07}, SciPy \citep{scipy}, seaborn \citep{seaborn}, pandas \citep{pandas}}


\appendix

\section{Full list of galaxies hosting CO/PAH outliers}\label{app:all}
\begin{figure*}
\centering
\includegraphics[width=\textwidth]{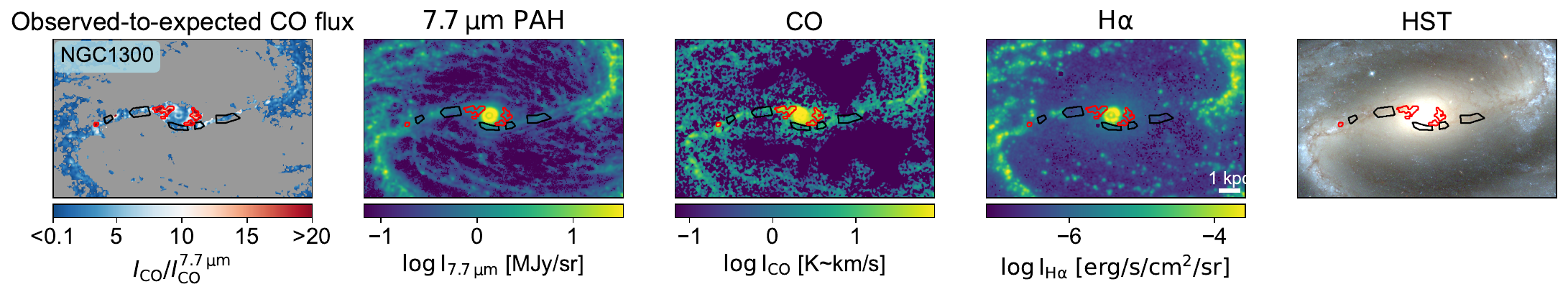}
\includegraphics[width=\textwidth]{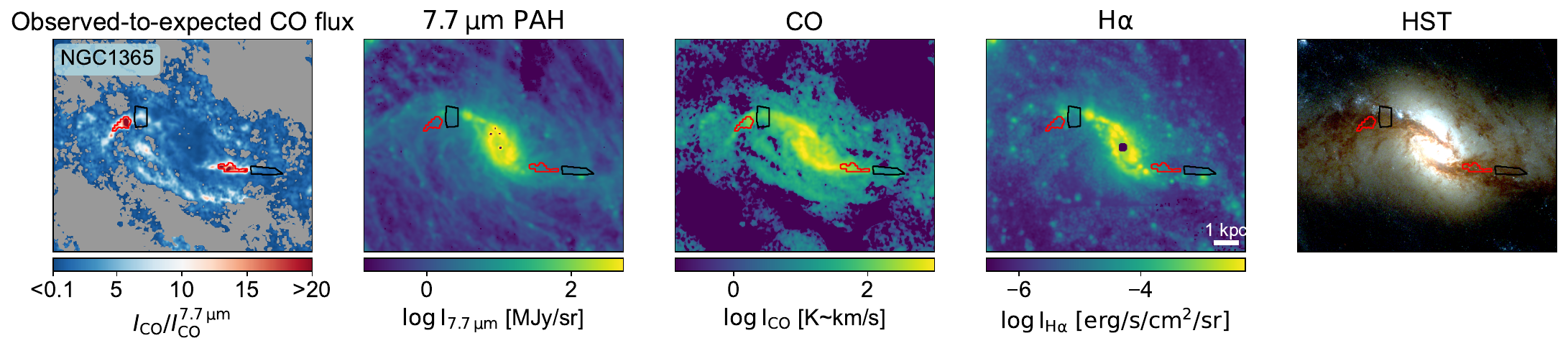}
\includegraphics[width=\textwidth]{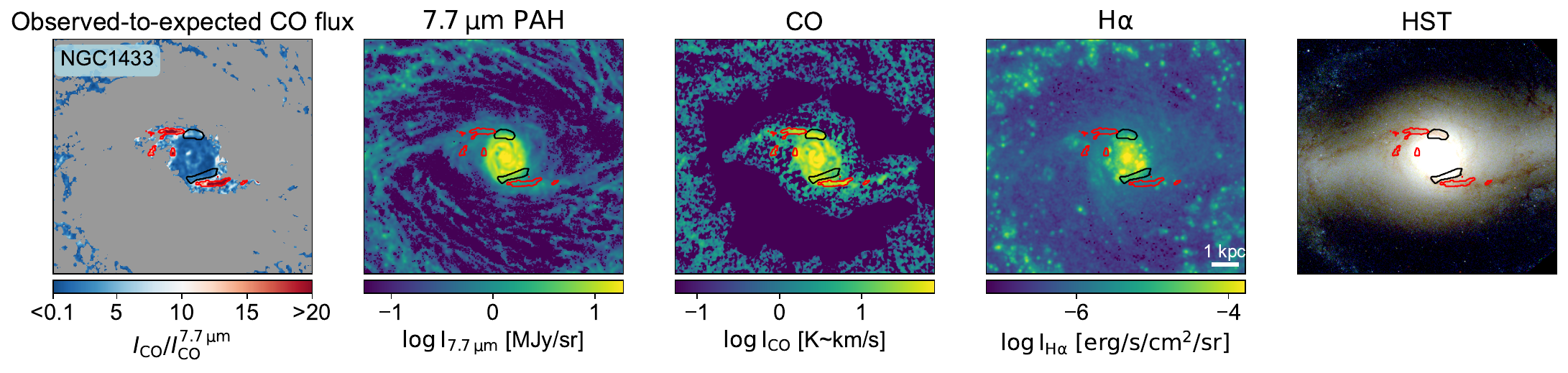}
\includegraphics[width=\textwidth]{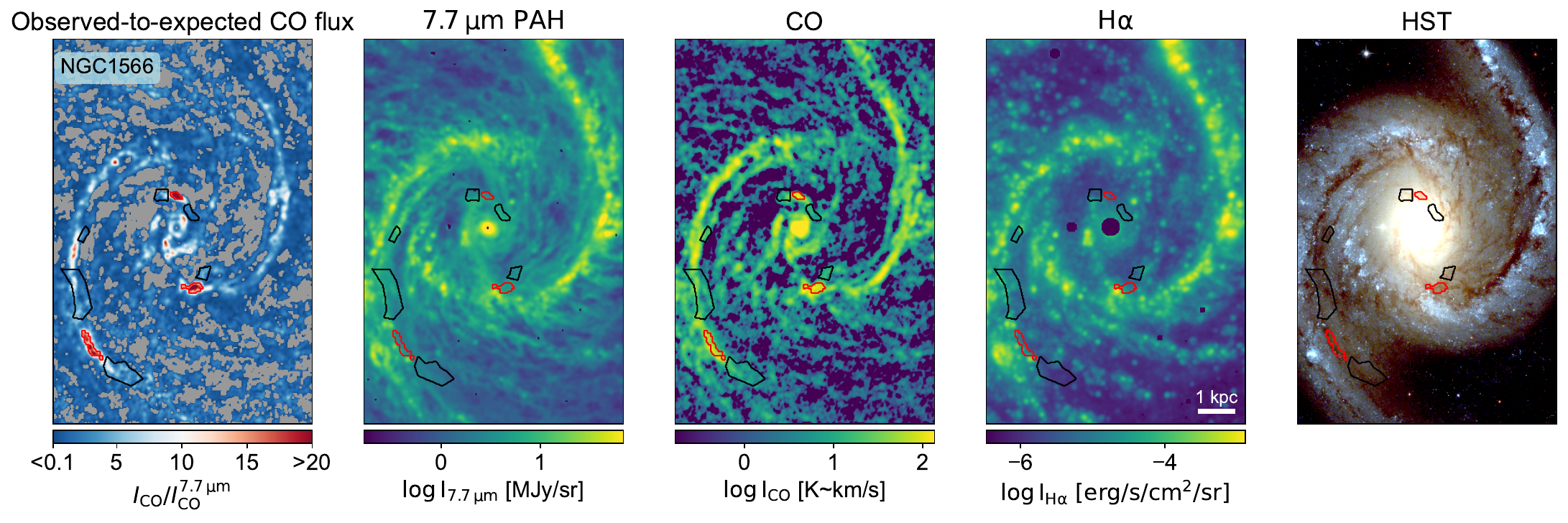}
\caption{JWST Cycle~1 galaxies (NGC\,1300, NGC\,1365, NGC\,1433, and NGC\,1566) hosting outlier regions. The first column shows the ratio of CO flux ($I_{\rm CO}$) to that expected from 7.7$\mu$m PAH emission ($I_{\mathrm{CO}}^{\mathrm{7.7\mu m}}$), using the relations from \citet{chown24}. Outlier regions, defined as having a flux ratio higher than 10 are highlighted in red, whereas null-test regions used in our analysis are shown in black. These outliers represent significant deviations from the CO-7.7\,$\mu$m PAH relation observed in \citet{chown24}. They are spatially localized in dynamically complex regions (bar lanes, spiral arms, and centers) and show no evidence of massive star formation in H$\alpha$. Subsequent columns show the 7.7$\mu$m PAH, CO, H$\alpha$, and HST observations (F814W, F555W, and F438W), where H$\alpha$ is from PHANGS-MUSE \citep{emsellem22} with extinction-correction \citep{belfiore22}.} \label{fig:fig6}
\end{figure*}

\begin{figure*}
\centering
\includegraphics[width=\textwidth]{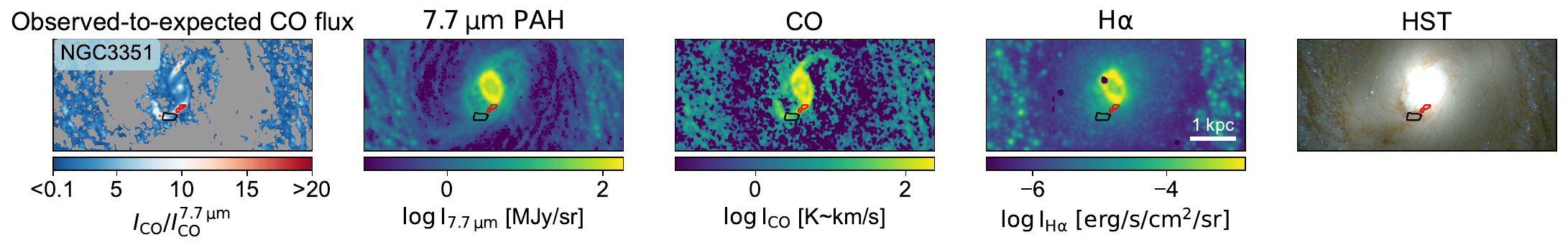}
\includegraphics[width=\textwidth]{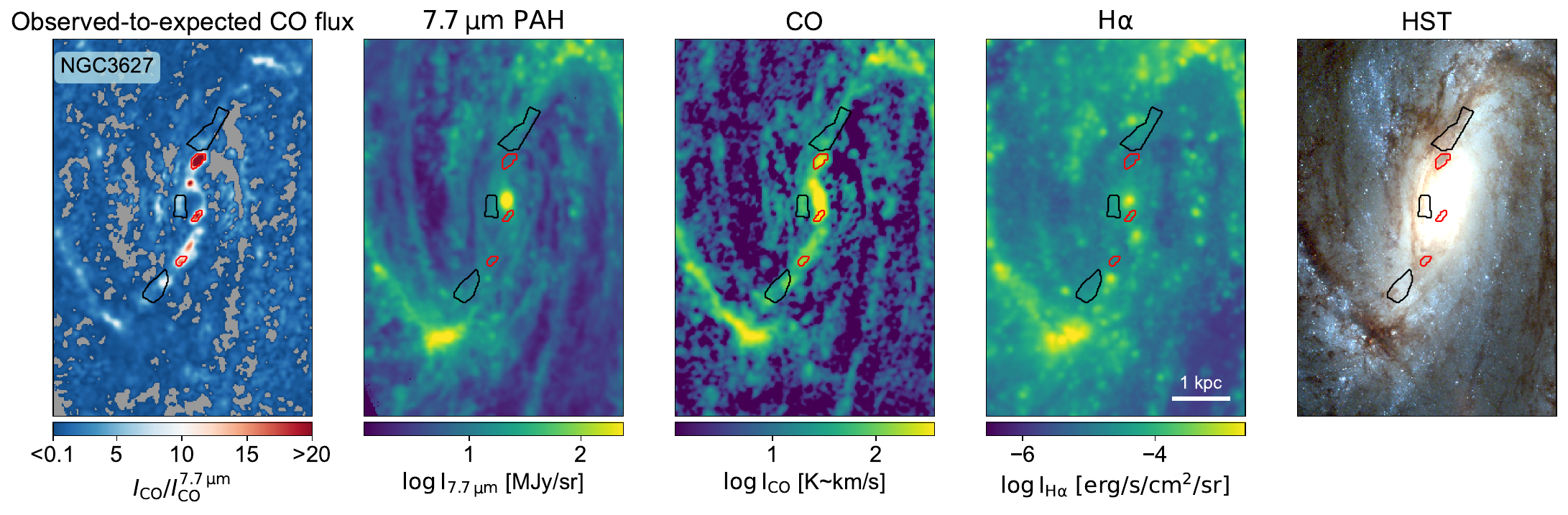}
\includegraphics[width=\textwidth]{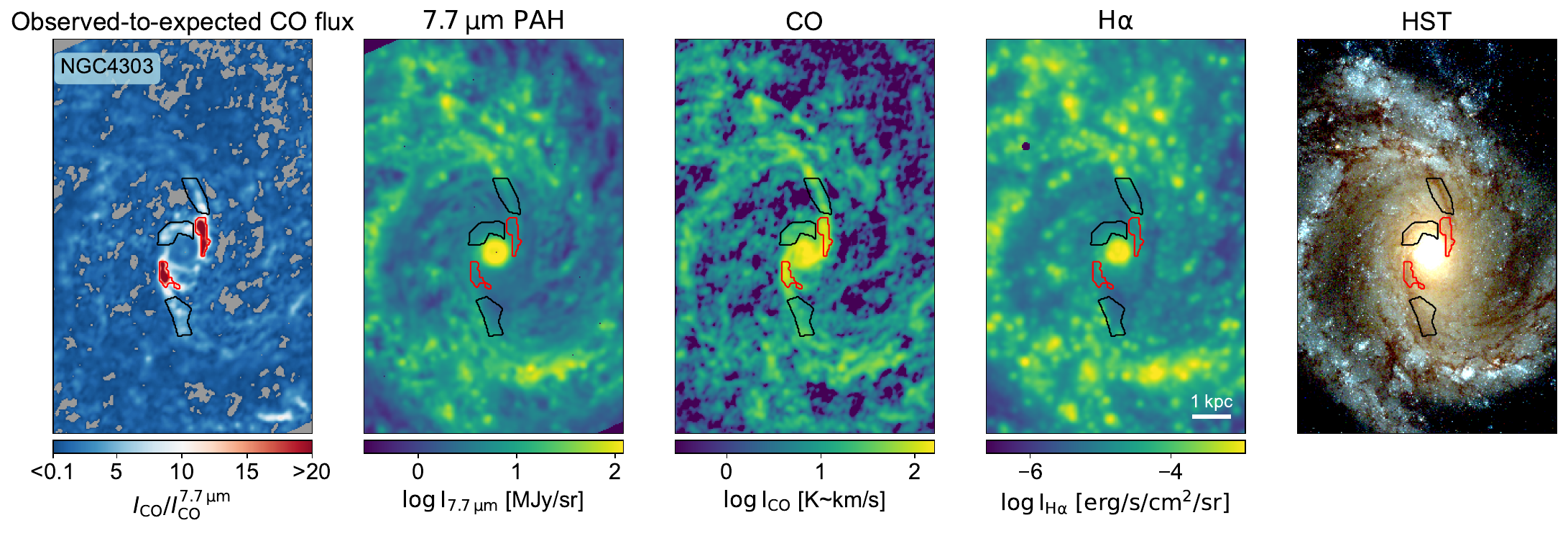}
\includegraphics[width=\textwidth]{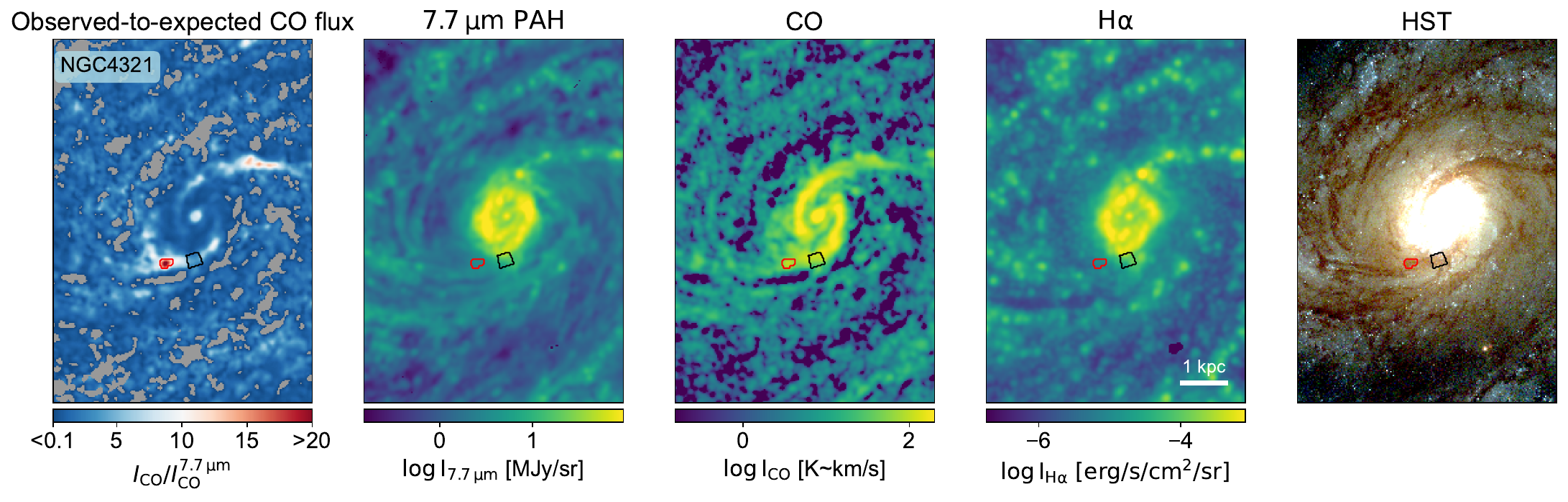}
\caption{Same as Figure~\ref{fig:fig6}, but for NGC\,3351, NGC\,3627, NGC\,4303, and NGC\,4321 which are part of JWST Cycle~1 galaxies hosting outlier regions.} \label{fig:fig7}
\end{figure*}

\begin{figure*}
\centering
\includegraphics[width=\textwidth]{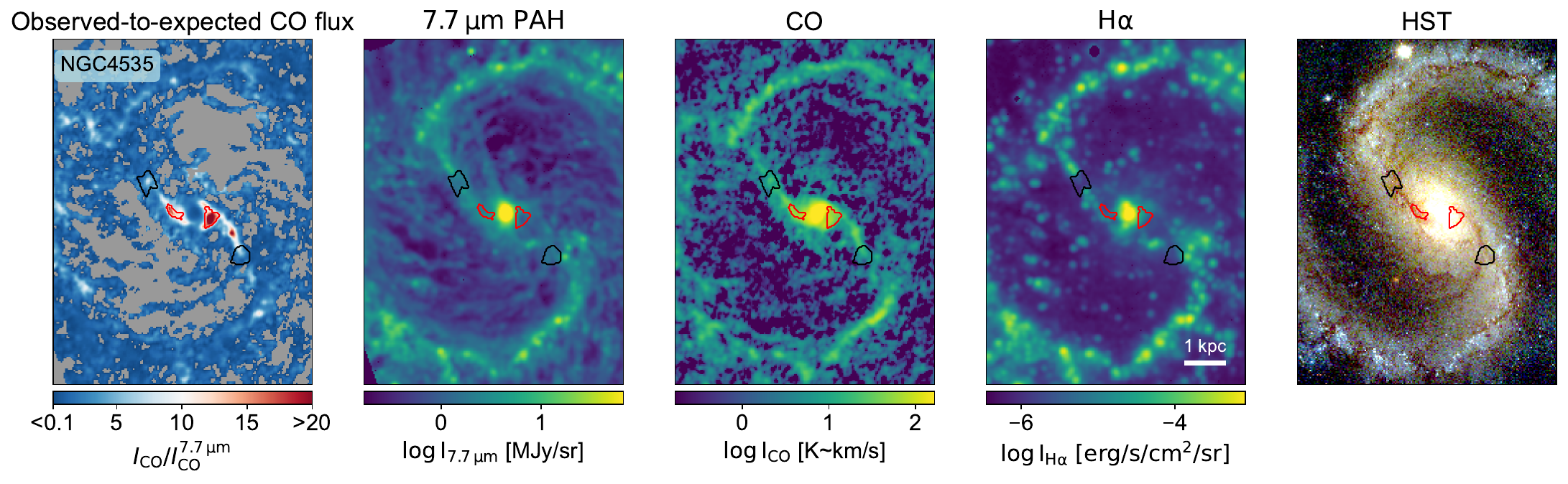}
\includegraphics[width=\textwidth]{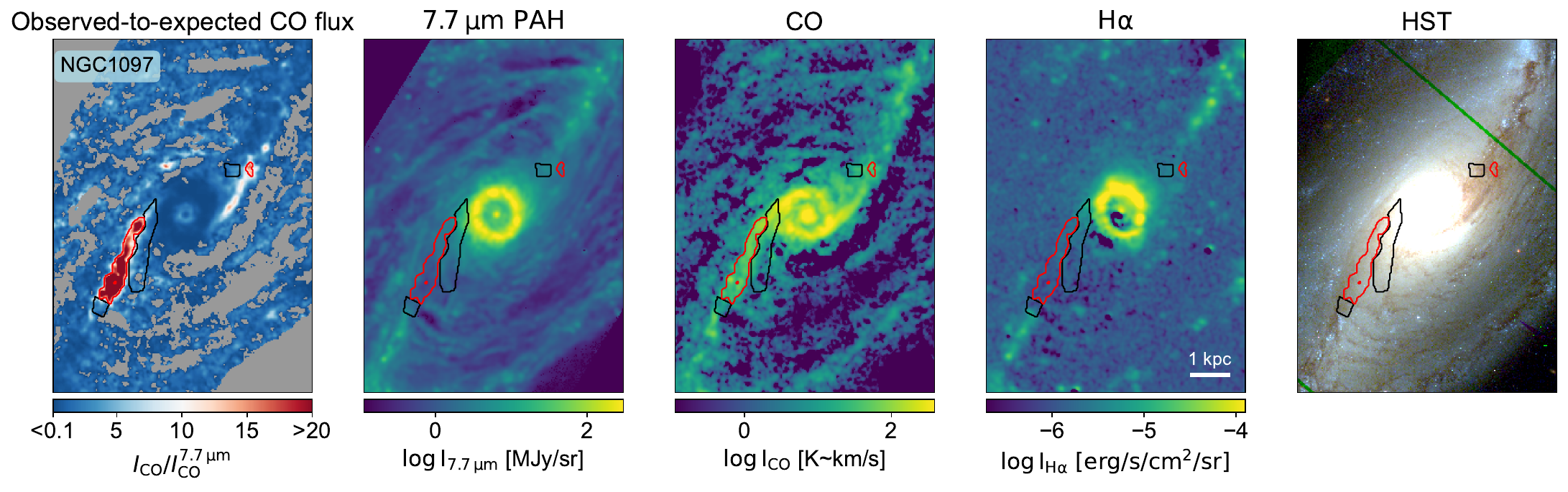}
\includegraphics[width=\textwidth]{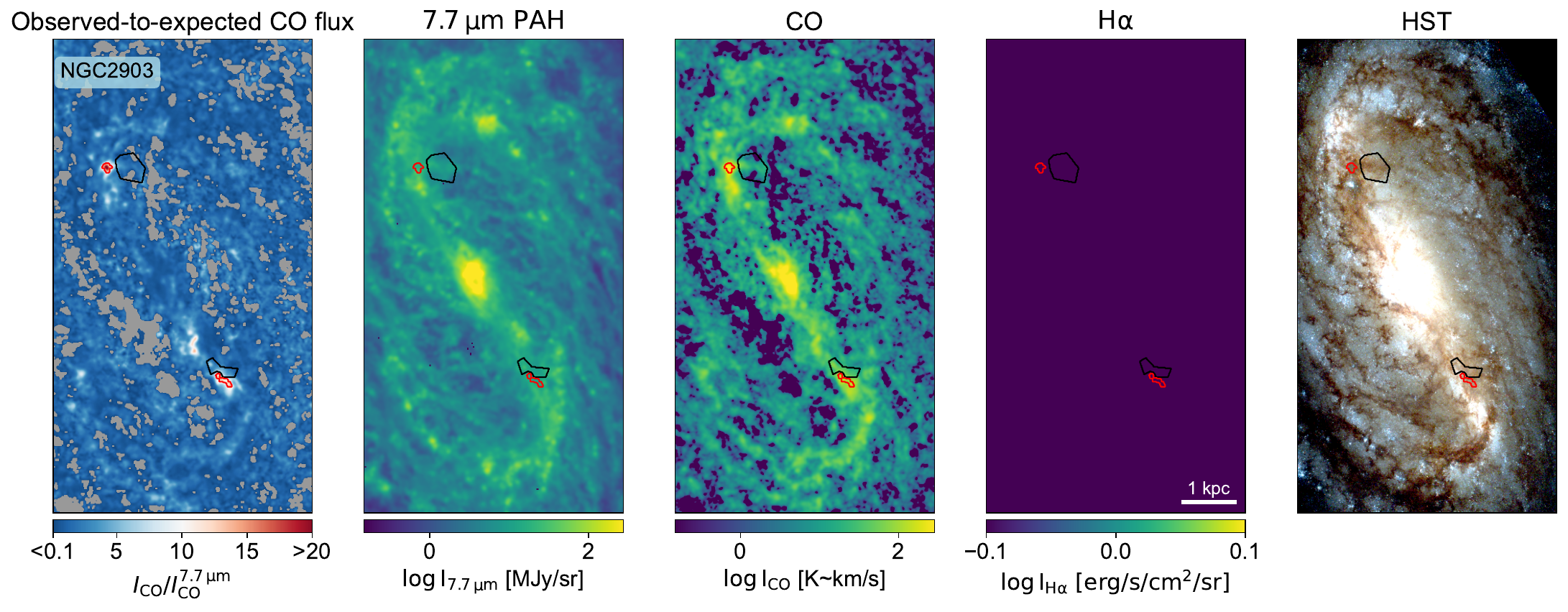}
\caption{Same as Figure~\ref{fig:fig6}, but for NGC\,4535 which is part of JWST Cycle~1 galaxies hosting outlier regions. NGC\,1097 and NGC\,2903 are from JWST Cycle~2 galaxies with HST data from \citet{lee22}. For Cycle~2 galaxies, H$\alpha$ is from narrowband imaging from Razza et al. (in preparation). Continuum subtracted H$\alpha$ map was not available for NGC\,2903 therefore not shown.} \label{fig:fig8}
\end{figure*}

\begin{figure*}
\centering
\includegraphics[width=\textwidth]{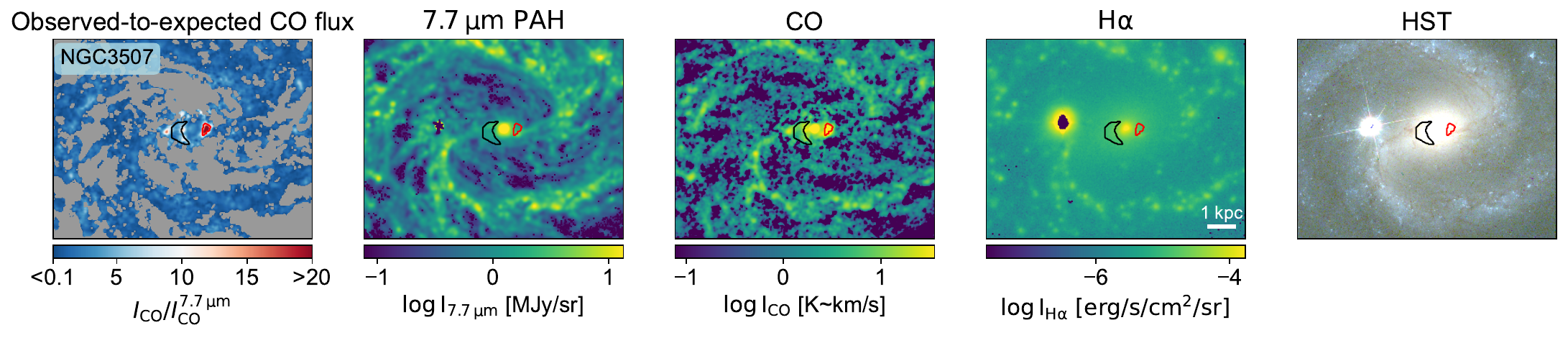}
\includegraphics[width=\textwidth]{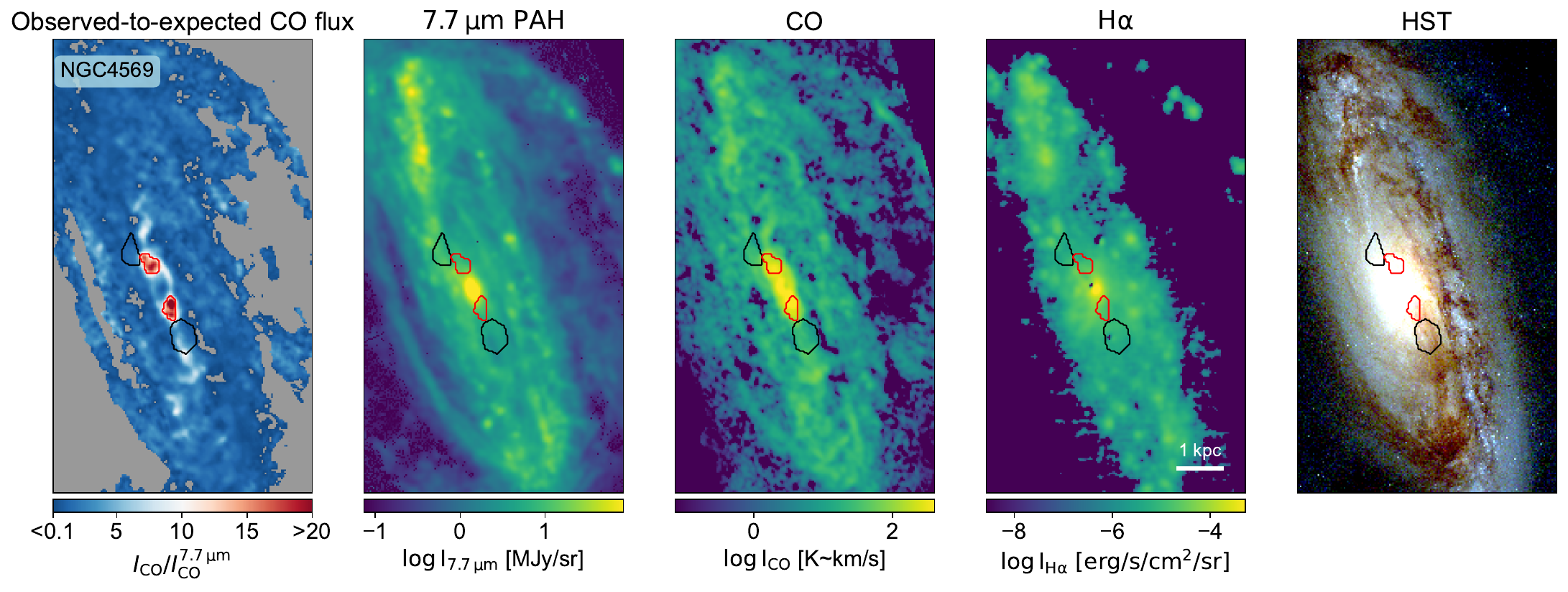}
\includegraphics[width=\textwidth]{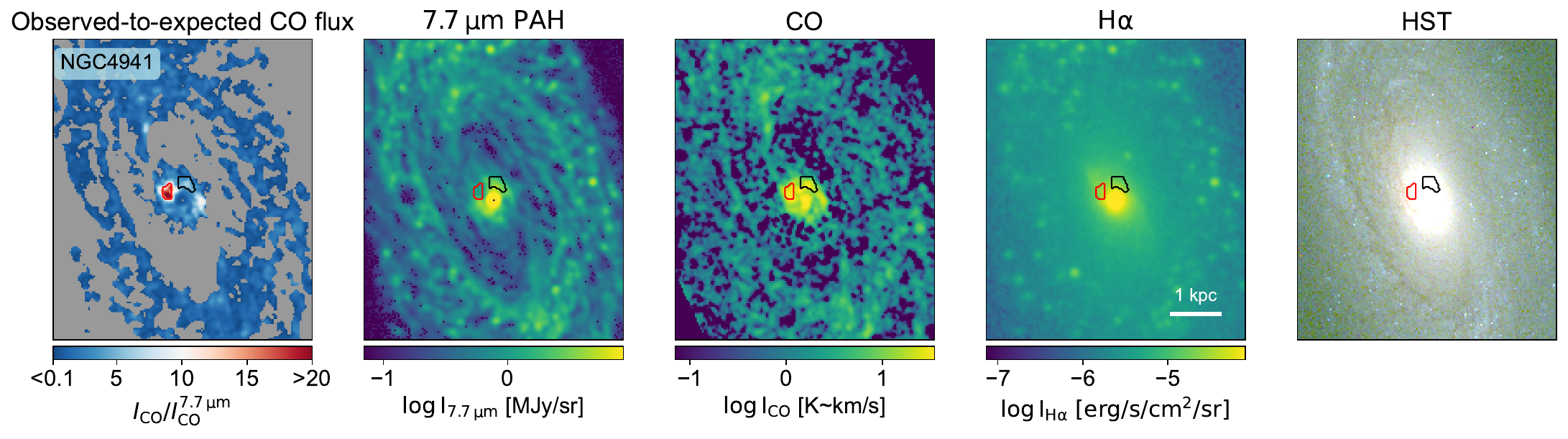}
\caption{Same as Figure~\ref{fig:fig6}, but for NGC\,3507, NGC\,4569, and NGC\,4941 from JWST Cycle~2 galaxies with HST data from \citet{lee22}. For Cycle~2 galaxies, H$\alpha$ is from narrowband imaging from Razza et al. (in preparation).} \label{fig:fig9}
\end{figure*}

Table~\ref{tab:prop} lists physical properties of 20 galaxies hosting outlier regions, together with the fraction of outlier pixels relative to the entire CO-emitting area. From Figures~\ref{fig:fig6} to \ref{fig:fig9}, we show all the galaxies studied in \citet{chown24} that host regions with abnormally high CO-to-PAH ratios. When available, three-color composite HST images from the PHANGS–HST survey \citep{lee22} are also included for some galaxies. Red contours highlight outlier regions, whereas their null-test regions are identified in black.

\begin{deluxetable*}{lcccccccc}
\tabletypesize{\scriptsize}
\tablewidth{0pt} 
\tablecaption{Summary of physical properties of galaxies hosting outliers \label{tab:prop}}
\tablehead{
\colhead{} & \colhead{} & \colhead{(a)} & \colhead{(b)} & \colhead{(c)} & \colhead{(d)} & \colhead{(e)} & \colhead{(f)} & \colhead{(g)}  \\
\colhead{Galaxy} &\colhead{JWST Cycle} & \colhead{$M_{\rm *}$} & \colhead{$\rm SFR$} & \colhead{$\rm sSFR$} &\colhead{Dist.} & \colhead{Incl.} & \colhead{Morphology} & \colhead{$f_{\rm{outlier}}$}\\
\colhead{} &\colhead{} &  \colhead{[$\rm log\,M_{\odot}$]} & \colhead{[$\rm log\,M_{\odot}/yr$]}& \colhead{[Gyr$^{-1}$]}  & \colhead{[Mpc]} & \colhead{[deg]}  & \colhead{type}  &\colhead{[\%]} \\
} 
\startdata 
NGC1300&Cycle 1&10.6&0.1&-1.5&18.99&31.8&Sbc&3.03\\
NGC1365&Cycle 1&11.0&1.2&-0.8&19.57&55.4&Sb&1.17\\
NGC1433&Cycle 1&10.9&0.1&-1.8&18.63&28.6&SBa&4.03\\
NGC1566&Cycle 1&10.8&0.7&-1.1&17.69&29.5&SABb&0.64\\
NGC3351&Cycle 1&10.4&0.1&-1.2&9.96&45.1&Sb&0.25\\
NGC3627&Cycle 1&10.8&0.6&-1.2&11.32&57.3&Sb&0.20\\
NGC4303&Cycle 1&10.5&0.7&-0.8&16.99&23.5&Sbc&0.59\\
NGC4321&Cycle 1&10.7&0.6&-1.2&15.21&38.5&SABb&0.06\\
NGC4535&Cycle 1&10.5&0.3&-1.2&15.77&44.7&Sc&0.53\\
\hline
NGC1097&Cycle 2&10.8&0.7&-1.1&13.58&48.6&SBb&2.04\\
NGC2566&Cycle 2&10.7&0.9&-0.8&23.44&48.5&Sb&7.43\\
NGC2903&Cycle 2&10.6&0.5&-1.1&10.0&66.8&Sbc&0.15\\
NGC2997&Cycle 2&10.7&0.6&-1.1&14.06&33.0&SABc&0.04\\
NGC3507&Cycle 2&10.4&-0.0&-1.4&23.55&21.7&SBb&0.34\\
NGC4569&Cycle 2&10.8&0.1&-1.7&15.76&70.0&Sab&0.70\\
NGC4579&Cycle 2&11.1&0.3&-1.8&21.0&40.22&Sb&1.08\\
NGC4731&Cycle 2&9.5&-0.2&-0.7&13.28&64.0&SBc&1.04\\
NGC4941&Cycle 2&10.2&-0.4&-1.5&15.0&53.4&SABa&0.35\\
NGC5643&Cycle 2&10.3&0.4&-0.9&12.68&29.9&Sc&1.18\\
NGC6300&Cycle 2&10.5&0.3&-1.2&11.58&49.6&SBb&1.84\\
\enddata
\tablecomments{(a) (b), and (c) -- Global stellar mass, star formation rate, and specific star formation rate from PHANGS--ALMA \citep{leroy19, leroy21_survey}. (d) and (e) -- Distance from \citet{anand21} and inclination from \citet{lang20}. (f) -- Galaxy morphology from \citet{paturel98}. (g) -- Fraction of outlier pixels relative to the entire CO-emitting region. }
\end{deluxetable*}

\begin{figure*}
\centering
\includegraphics[width=\textwidth]{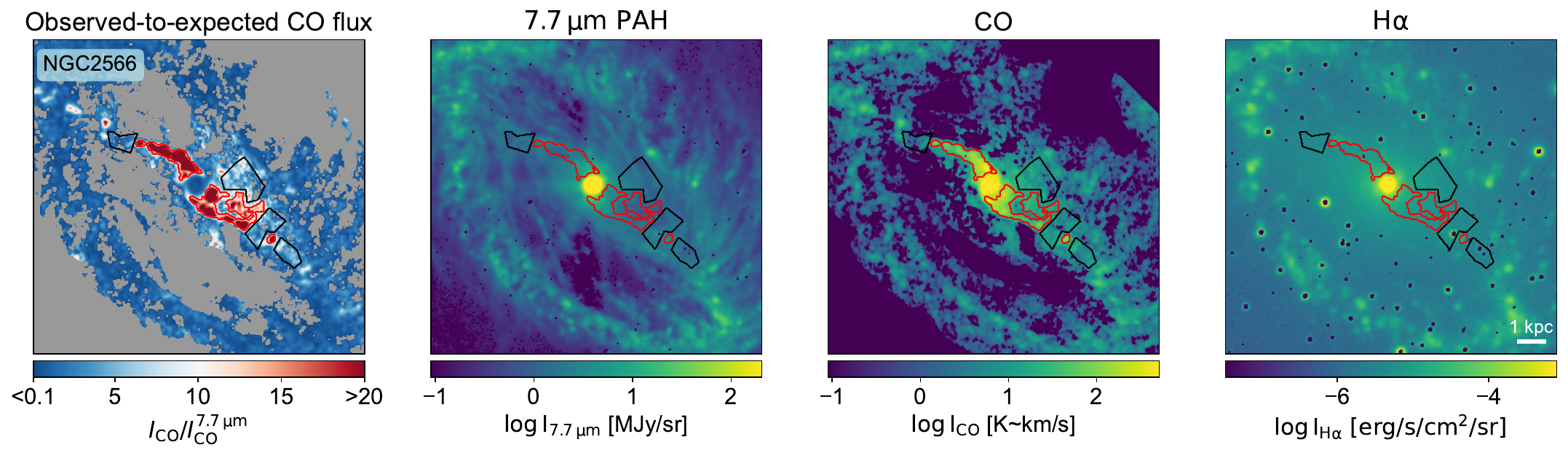}
\includegraphics[width=\textwidth]{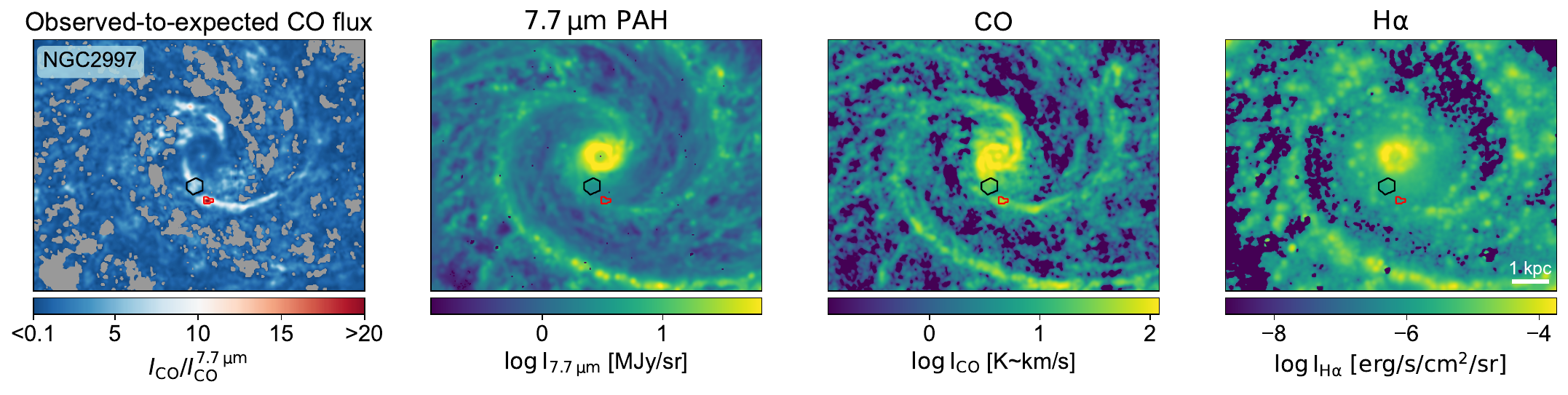}
\includegraphics[width=\textwidth]{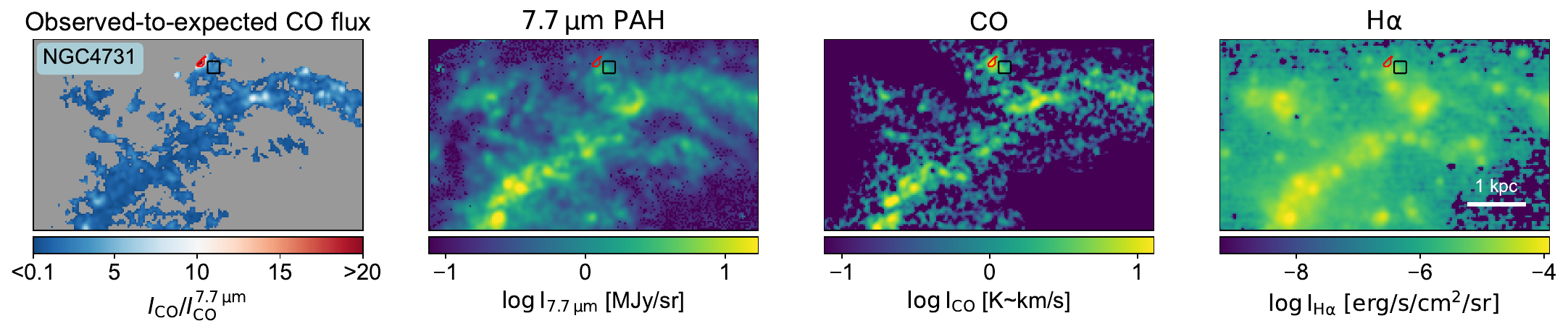}

\caption{Same as Figure~\ref{fig:fig6}, but for NGC\,2226, NGC\,2997, and NGC\,4731 from JWST Cycle~2 galaxies with HST data from \citet{lee22}. H$\alpha$ is from narrowband imaging from Razza et al. (in preparation).} \label{fig:fig10}
\end{figure*}

\begin{figure*}
\centering
\includegraphics[width=\textwidth]{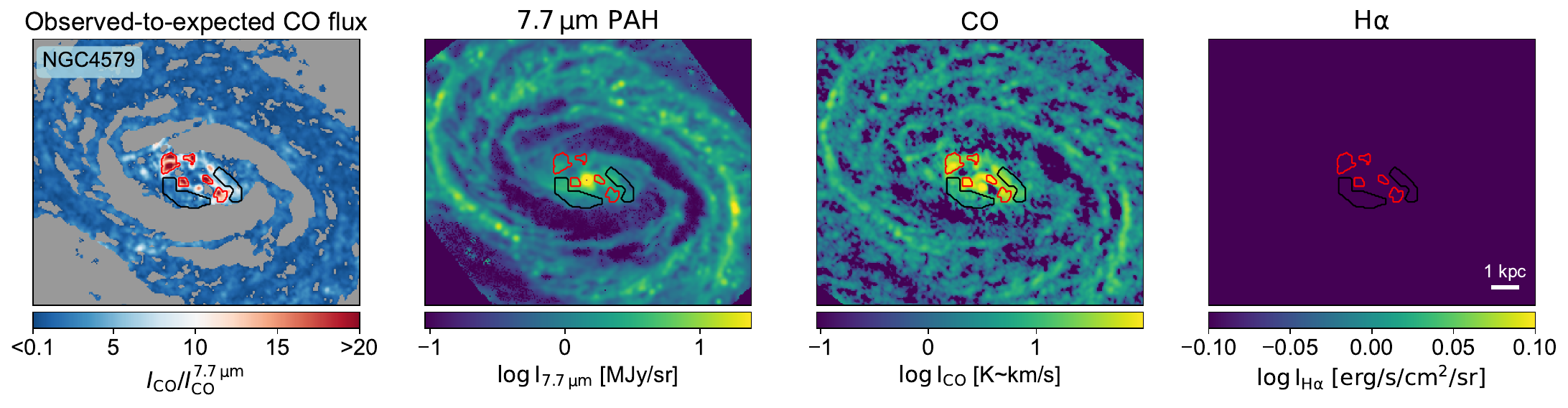}
\includegraphics[width=\textwidth]{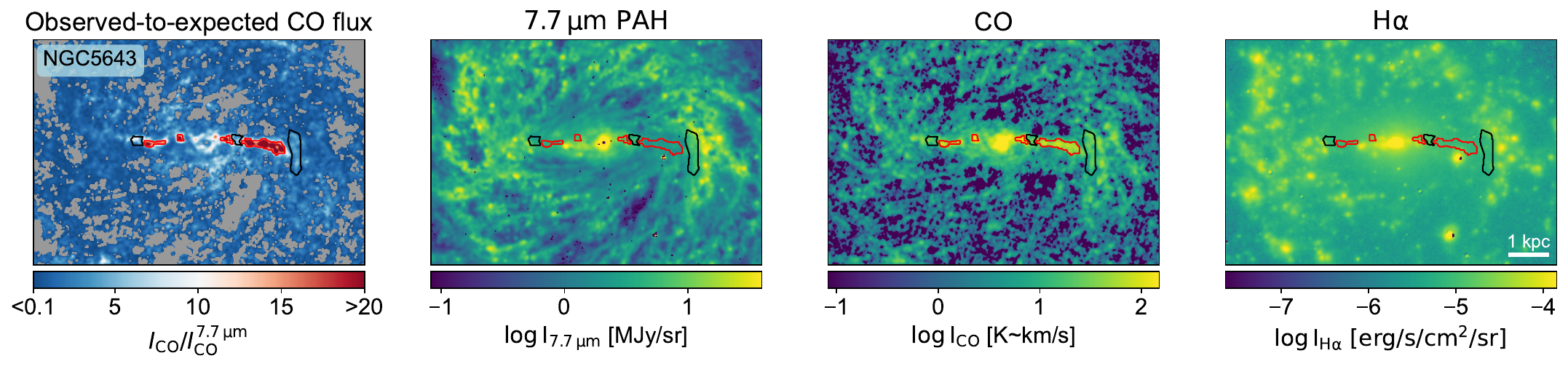}
\includegraphics[width=\textwidth]{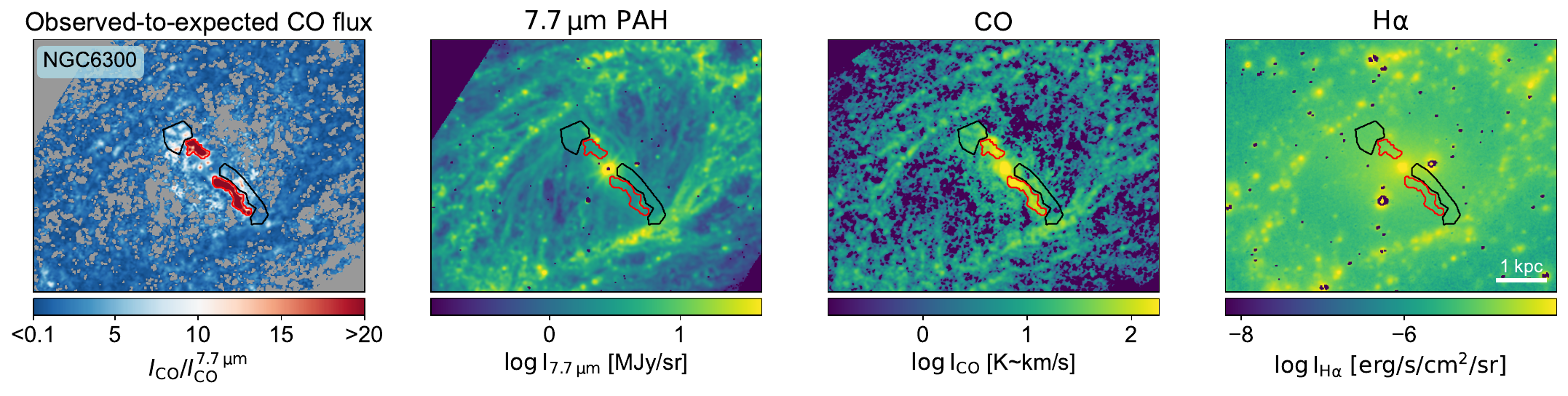}
\caption{Same as Figure~\ref{fig:fig6}, but for the rest of JWST Cycle~2 galaxies where H$\alpha$ is from narrowband imaging (Razza et al. in preparation). Continuum subtracted H$\alpha$ map is not available for NGC\,4579 therefore not shown.} \label{fig:fig11}
\end{figure*}

\section{CO velocity dispersions in individual galaxies}\label{app:vdisp_c2}
\begin{figure*}
\centering
\includegraphics[width=\textwidth]{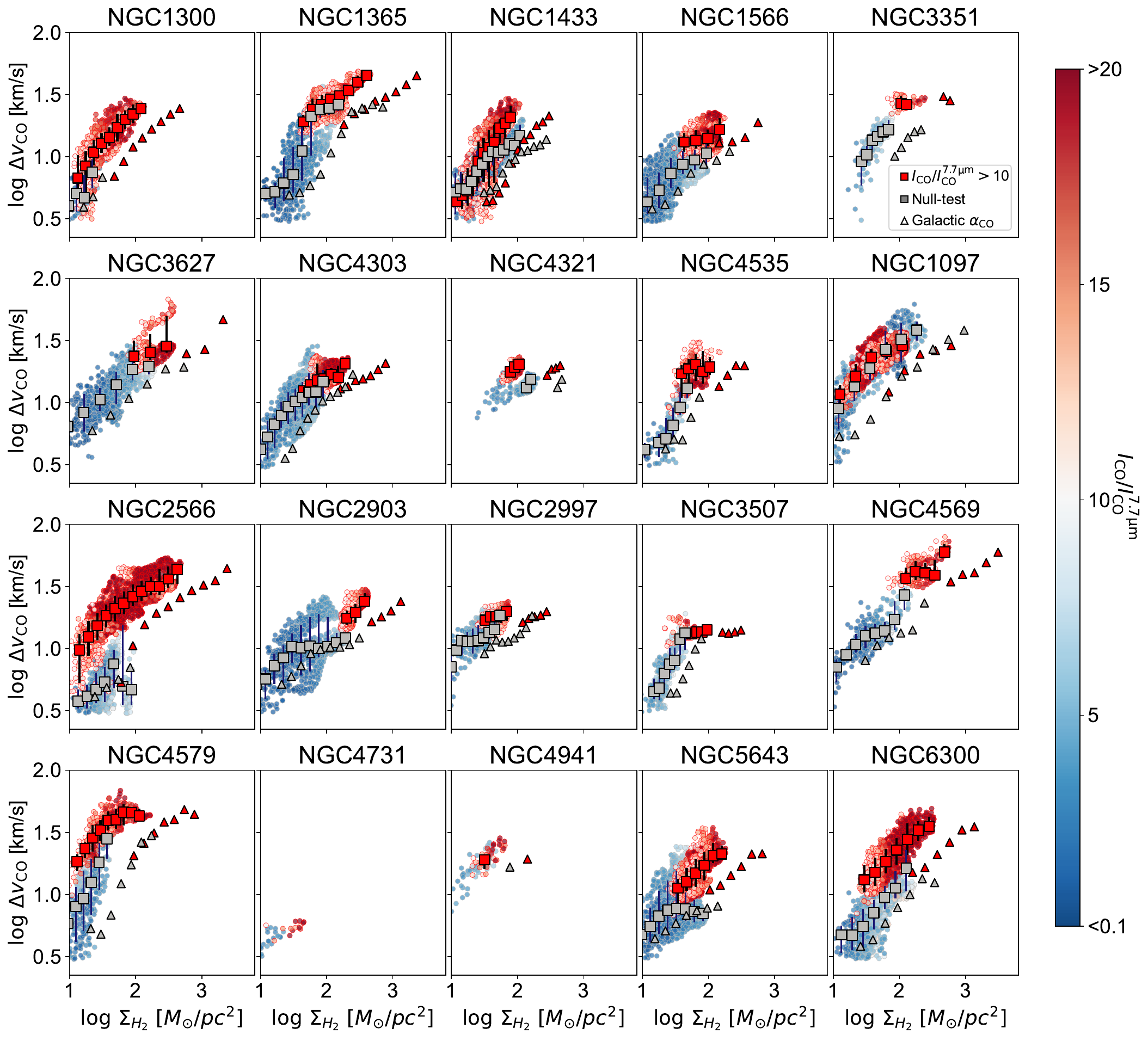}
\caption{Similar to the bottom panel of Figure~\ref{fig:vdisp_all}, CO velocity dispersion ($\Delta v_{\mathrm{CO}}$) is shown as a function of molecular gas surface density ($\Sigma_{\mathrm{H_{2}}}$) for all 20 galaxies hosting outlier regions. Data points in circle are color coded by the observed-to-expected CO flux ratio ($I_{\mathrm{CO}}/I_{\mathrm{CO}}^{\mathrm{7.7\mu m}}$), where points corresponding to outlier regions are outlined in red. We show the median $\Delta v_{\mathrm{CO}}$ in logarithmically spaced bins of $\log \Sigma_{\mathrm{H_{2}}}$ together with 1$\sigma$ scatter for each outlier and null test region (red and gray squares, respectively). While $\Sigma_{\mathrm{H_{2}}}$ is estimated using the $\alpha_{\mathrm{CO}}$–$\Delta v_{\mathrm{CO}}$ relation from \citet{teng24}, we also show the median of $\Delta v_{\mathrm{CO}}$ as a function of $\Sigma_{\mathrm{H_{2}}}$ when a constant galactic $\alpha_{\mathrm{CO}}=\Sigma_{\rm H_2}/I_{\rm CO}=4.35 \mathrm{M_{\odot}pc^{-2}/(K~km~s^{-1})}$ is adopted instead (triangles).} \label{fig:fig10_vdisp}
\end{figure*}

Figure~\ref{fig:fig10_vdisp} compares CO velocity dispersion ($\Delta v_{\rm{CO}}$) as a function of molecular gas surface density ($\Sigma_{\mathrm{H_{2}}}$) between outlier and null-test regions for all 20 galaxies hosting outlier regions.

\bibliographystyle{aasjournalv7}
\bibliography{mybib}
\suppressAffiliationsfalse
\allauthors


\end{document}